\documentclass[final,5p,times,twocolumn,authoryear]{elsarticle}

\usepackage{amssymb}
\usepackage{amsmath}
\usepackage{graphicx}
\usepackage{float}
\usepackage{subfig}
\usepackage{multirow}
\usepackage{booktabs}
\usepackage{xcolor}
\usepackage{newunicodechar}
\newunicodechar{≈}{\approx}
\usepackage{placeins}

\usepackage[
  colorlinks=true,
  citecolor=blue,
  linkcolor=black,
  urlcolor=blue
]{hyperref}

\journal{Nuclear Instruments and Methods in Physics Research Section A}

\begin{document}

\begin{frontmatter}

\title{Ion Backflow Suppression and Absorption Dynamics in Cascaded GEM Structures}

\author[1]{Md Kaosor Ali Mondal\corref{cor1}}
\ead{mohammad.mondal@cern.ch}

\author[1]{Shakir Eqbal}
\ead{di2505@students.iitmandi.ac.in}

\author[1]{Sachin Rana}
\ead{sachin.rana@cern.ch}

\author[1]{Poojan Angiras}
\ead{poojan.angiras@cern.ch}

\author[1]{A. Sarkar}
\ead{amal.sarkar@cern.ch}

\cortext[cor1]{Corresponding author}

\affiliation[1]{organization={Indian Institute of Technology Mandi},
            city={Mandi},
            postcode={175005},
            state={Himachal Pradesh},
            country={India}}

\begin{abstract}
Ion backflow (IBF) is a persistent challenge in Gas Electron Multiplier (GEM) detectors:
positive ions drifting back into the sensitive volume distort the drift field and degrade
both spatial and energy resolution. We investigate the ion-absorption characteristics of
single- to quintuple-GEM detectors using ANSYS and Garfield$^{++}$ for an
Ar/CO$_2$ (70:30) gas mixture. A layer-resolved formalism is developed that
distinguishes conditional foil-level absorption from a detector-level absorption metric
that correctly weights ion production at every amplification stage. For the conventional
biconical geometry, cascading foils improves the overall ion-absorption fraction only
moderately, from 47.83\% for a single GEM to 59.06\% for a quintuple GEM. Replacing
the biconical holes with a single-conical profile raises the absorption fraction to
75\%-87\% across the same range of multiplicities and reduces the IBF/Gain
ratio by a factor of two to three an improvement that exceeds what is gained by adding
four additional biconical foils. As a secondary test under more realistic conditions, the
second transfer-gap length of a triple-GEM detector is varied from 1.2 to 3.5~mm. The
gain shows a non-monotonic dependence with a maximum near 2.0--2.5~mm, while the
IBF/Gain ratio decreases monotonically with increasing gap length; the single-conical
geometry retains its factor-of-two advantage throughout. These findings establish that
ion suppression in cascaded GEM structures is governed not only by the number of foils
but primarily by the local ion-collection properties of each foil, and that single-conical
holes represent a promising route to improved performance in high-rate detector applications.
\end{abstract}


\end{frontmatter}

\section{Introduction}
\label{sec:introduction}

Gaseous detectors have been at the heart of high-energy physics experiments for decades.
The invention of the Multiwire Proportional Chamber (MWPC) by Georges Charpak at CERN in
1968~\cite{charpak1970some,charpak1979multiwire,sauli1977principles} marked a landmark advance,
raising the rate capability of gaseous detectors by orders of magnitude and enabling their
widespread adoption across a broad range of experiments~\cite{grupen2008particle,tsyganov2008gas}.
Despite this success, the MWPC suffers from a fundamental limitation: positive ions produced
during gas multiplication accumulate in the drift region, distorting the electric field and
degrading detector performance. The subsequent introduction of the Micro-Strip Gas Counter
(MSGC) by Anton Oed in 1988~\cite{oed1995properties} improved position resolution and rate
capability but did not eliminate this ion-backflow problem. These limitations motivated the
development of Micro-Pattern Gaseous Detectors (MPGDs~\cite{titov2013micro,shekhtman2002micro}),
a family of devices that confine gas multiplication to micron-scale structures.

Among MPGDs, the Gas Electron Multiplier (GEM), introduced by F.~Sauli at CERN in
1997~\cite{sauli1997gem} and further developed in
Refs.~\cite{sauli2004progress,benlloch1998development,sauli2003development}, has become
the most widely deployed
architecture~\cite{sauli2016gem,ketzer2004performance,CMS:2015GEMTDR,ALICETPC:2021upgrade}.
A GEM foil consists of a thin polyimide substrate, clad on both sides with copper and
perforated by a dense array of micron-scale holes. When a high voltage is applied across the
foil, strong electric fields develop inside the holes and drive avalanche multiplication of
drifting primary electrons. A large fraction of the resulting positive ions is captured on
the copper electrodes of the foil itself, substantially reducing the ion flux that would
otherwise return to the drift
region~\cite{bachmann1999charge,sauli2003electron}.

Ion backflow, defined as the fraction of avalanche ions that escape the amplification region
and drift back toward the cathode, is of central importance for detectors operating at high
particle rates. Back-drifting ions accumulate into a quasi-static positive-space-charge
distribution that distorts the drift field, introduces position-dependent track-reconstruction
errors, and ultimately limits the rate capability of large-volume devices such as Time
Projection Chambers (TPCs)~\cite{bondar2003study,boehmer2013simulation,ALICE:2013TPCTDR,ALICE:2015TPCTDRadd}.
Effective ion suppression is therefore critical for maintaining gain stability, field
uniformity, and long-term operational reliability.

Cascading multiple GEM foils is the standard approach to IBF suppression: each upstream foil presents an additional opportunity for ions to terminate on an electrode rather than continue drifting toward the cathode. It has further been established that changing the GEM hole profile from the conventional biconical shape to a single-conical geometry increases the effective gain and reduces ion backflow~\cite{lyashenko2006advances,lyashenko2007further}. The single-conical profile has the additional practical advantage that it can be produced with a single photolithographic mask, eliminating the alignment step required for double-sided biconical etching and thereby simplifying fabrication~\cite{villa2011progress}.

Despite these observations, a systematic, layer-resolved study of how ion absorption evolves with GEM multiplicity in the single-conical geometry and a detector-level metric that correctly accounts for ion production at every amplification stage throughout the cascade has not previously been reported. The present work addresses this gap. We study ion absorption for single- to quintuple-GEM detectors with both the conventional biconical and the modified single-conical hole profiles, develop a rigorous layer-resolved formalism that separately quantifies the conditional and absolute absorption contributions of each foil, and introduce a weighted detector-level absorption measure. As a secondary validation, we examine how the second transfer-gap length of a triple-GEM detector influences gain, IBF, and the IBF/Gain ratio for both geometries, testing whether the advantage of the single-conical profile persists under a more realistic, non-uniform gap configuration.

The central finding is that the ion-suppression capability of cascaded GEM structures is
governed not only by the number of foils but also and more strongly by the local
ion-collection properties of each foil. Switching from biconical to single-conical holes
raises the overall ion-absorption fraction by approximately 25\% points across all
multiplicities studied, an improvement that exceeds what can be achieved by adding four
further biconical foils. The transfer-gap study confirms that this advantage is robust under
realistic detector geometries.

\section{Methodology}
\label{sec:methodology}

\begin{figure}[!t]
  \centering
  \includegraphics[width=\columnwidth]{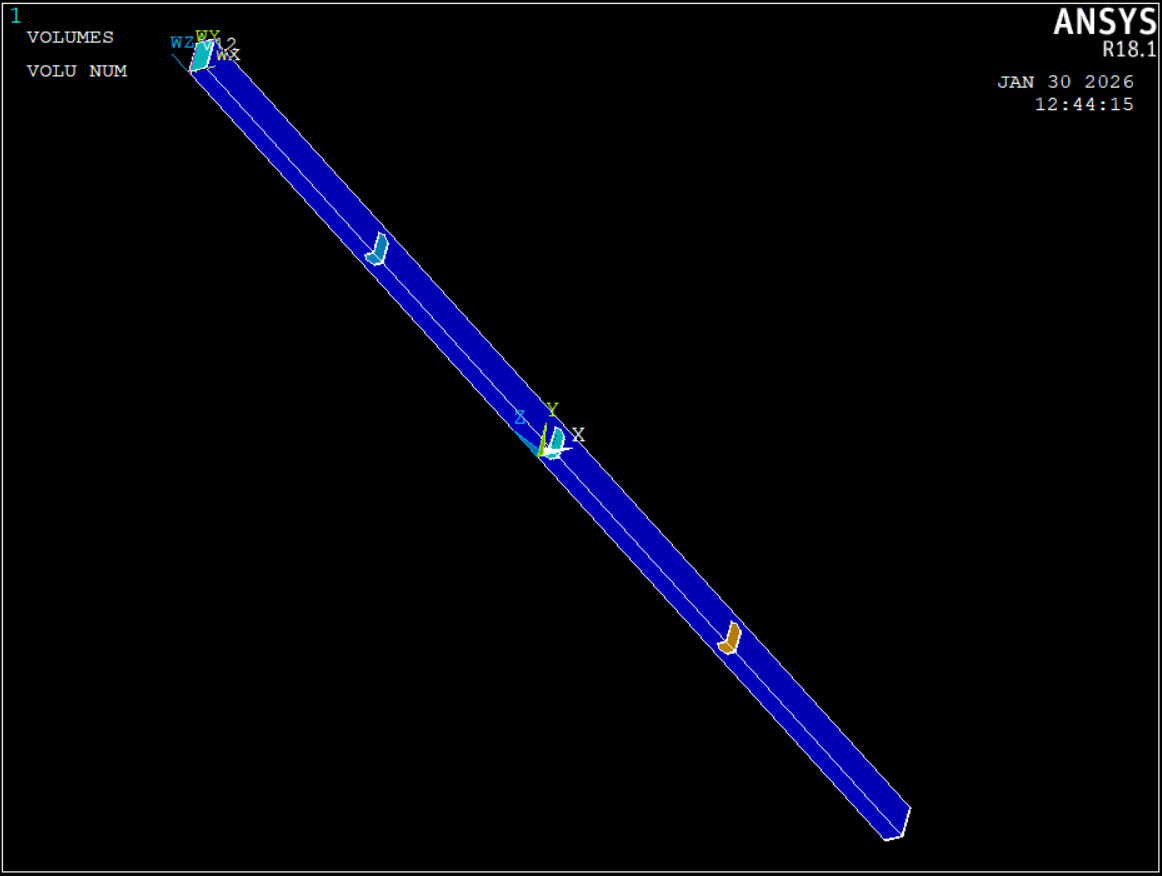}
  \caption{(Color online) Triple-GEM geometry constructed in ANSYS Mechanical APDL,
  with three equally spaced foils.}
  \label{fig:TripleGEM_APDL}
\end{figure}

Ion absorption was investigated using a two-stage computational workflow. GEM geometries
were first constructed in ANSYS~\cite{martyanov2014ansys,thompson2017ansys} to obtain
accurate three-dimensional electrostatic field maps, which were then imported into
Garfield$^{++}$~\cite{Garfieldppusemanual,veenhof1998garfield} for microscopic
particle tracking. A representative triple-GEM geometry, as built in ANSYS Mechanical
APDL, is shown in Figure~\ref{fig:TripleGEM_APDL}, and the full set of simulation
parameters is given in Table~\ref{tab:sim_parameters}.

\begin{table}[t]
\centering
\caption{Simulation parameters. Unless stated otherwise, values refer to the main study
of Section~\ref{sec:results}.}
\label{tab:sim_parameters}
\resizebox{\columnwidth}{!}{%
\begin{tabular}{ll}
\toprule
Parameter & Value \\
\midrule
Gas mixture & Ar/CO$_2$ (70:30) \\
Kapton thickness & 50~$\mu$m \\
Cu thickness & 5~$\mu$m (each side) \\
Inner (biconical) / upper (single-conical) hole diameter & 50~$\mu$m \\
Outer (biconical) / lower (single-conical) hole diameter & 70~$\mu$m \\
Pitch & 140~$\mu$m \\
GEM voltage, $\Delta V_{\mathrm{GEM}}$ & 300~V \\
Drift gap & 1~mm \\
Transfer gap & 1~mm (main study) \\
Second transfer gap (Section~\ref{sec:transfer_gap}) & 1.2--3.5~mm \\
Induction gap & 1~mm \\
Hole field & $\approx$40--45~kV/cm \\
Drift/transfer/induction fields & $\approx$2--3~kV/cm \\
Number of GEM foils & 1--5 \\
Field solver & ANSYS (Maxwell, Mechanical APDL) \\
Transport model & Garfield$^{++}$ microscopic (\texttt{AvalancheMicroscopic}) \\
Gas transport & Magboltz \\
Primary electrons per run ($N_{\rm primary}$) & 2000 \\
\bottomrule
\end{tabular}%
}
\end{table}

GEM geometries for single- to quintuple-stack configurations were designed in ANSYS Maxwell.
Applied potentials were tuned to maintain uniform field strengths in each region: approximately
40--45~kV/cm inside the GEM holes to drive avalanche multiplication, and 2--3~kV/cm in the
drift, transfer, and induction gaps. All gap lengths were kept equal at 1~mm in the main study
to ensure fair comparison across multiplicities. The resulting electrostatic solutions were
exported as field-map files using the ComponentAnsys123 interface and imported into
Garfield$^{++}$.

Gas transport properties for Ar/CO$_2$ (70:30) were computed with Magboltz~\cite{biagi1999monte},
providing drift velocities, diffusion coefficients, and Townsend coefficients. Electron
avalanches were then simulated with the \texttt{AvalancheMicroscopic} class, which tracks
individual charge carriers through a full Monte Carlo procedure accounting for ionization,
attachment, and gain fluctuations. Ion trajectories and endpoints were recorded throughout
each simulation, enabling quantitative assessment of absorption fractions and ion backflow
across all detector configurations.

Each detector configuration was simulated with an ensemble of $N_{\rm primary} = 2000$ primary
electrons to ensure adequate statistical precision. For the ion-absorption fractions and
the IBF/Gain ratio, the statistical uncertainties were estimated from the Poisson
fluctuations in the tracked electron and ion counts: $\sigma = \sqrt{N}/N_{\rm total}$,
where $N$ is the number of absorbed (or back-drifting) particles and $N_{\rm total}$ is
the total ion population. These uncertainties are propagated in quadrature where relevant
and are shown as error bars in Figures~\ref{fig:GainVsT2}--\ref{fig:RatioVsT2}. For the
main absorption study (Sections~\ref{sec:biconical} and~\ref{sec:single_conical}), the
per-foil contributions $A_i$ are quoted to the precision with which they can be read from
the simulated distributions; no error bars are shown in Figures~\ref{fig:biconical_ion_absorption}
and~\ref{fig:singleconical_ion_absorption} because the statistical fluctuations at
$N_{\rm primary} = 2000$ are below 0.5\% points for every foil and configuration
studied.

\section{Results}
\label{sec:results}

\subsection{Validation Against Experimental Data}
\label{sec:validation}

Before proceeding to parametric studies, we validate the ANSYS--Garfield$^{++}$ framework
against the experimental gain measurements of Bachmann et al.~\cite{bachmann1999charge},
who determined the effective gain of a single biconical GEM foil as a function of the
applied voltage $\Delta V_{\mathrm{GEM}}$, with the drift, induction, and intra-hole fields
held at 1.5~kV/cm, 5~kV/cm, and 50--60~kV/cm, respectively.

\subsubsection{Experimental Gain}

In those measurements the effective gain $M$ was extracted from the ratio of the anode
signal current $I_S$ to the primary ionization current. For a 6~keV quasi-monochromatic
X-ray beam at counting rate $R$, each absorbed photon produces on average $n \approx 220$
primary electron--ion pairs, giving a primary charge rate of $enR$. The gain is therefore
\begin{equation}
M = \frac{I_S}{e\,n\,R},
\label{eq:gain_def}
\end{equation}
where $e$ is the elementary charge.

\subsubsection{Simulation Setup and Comparison}

To replicate this measurement, the identical GEM geometry and field configuration were
implemented in ANSYS, and the resulting field maps were imported into Garfield$^{++}$.
Electron transport was then simulated starting from 220 electrons at an initial energy
of 27~eV, corresponding to the average energy per primary electron from a 6~keV photon
interaction. The effective gain was determined by counting the total number of electrons
produced by avalanche multiplication per primary transport electron.

\begin{figure}[!t]
  \centering
  \includegraphics[width=\columnwidth]{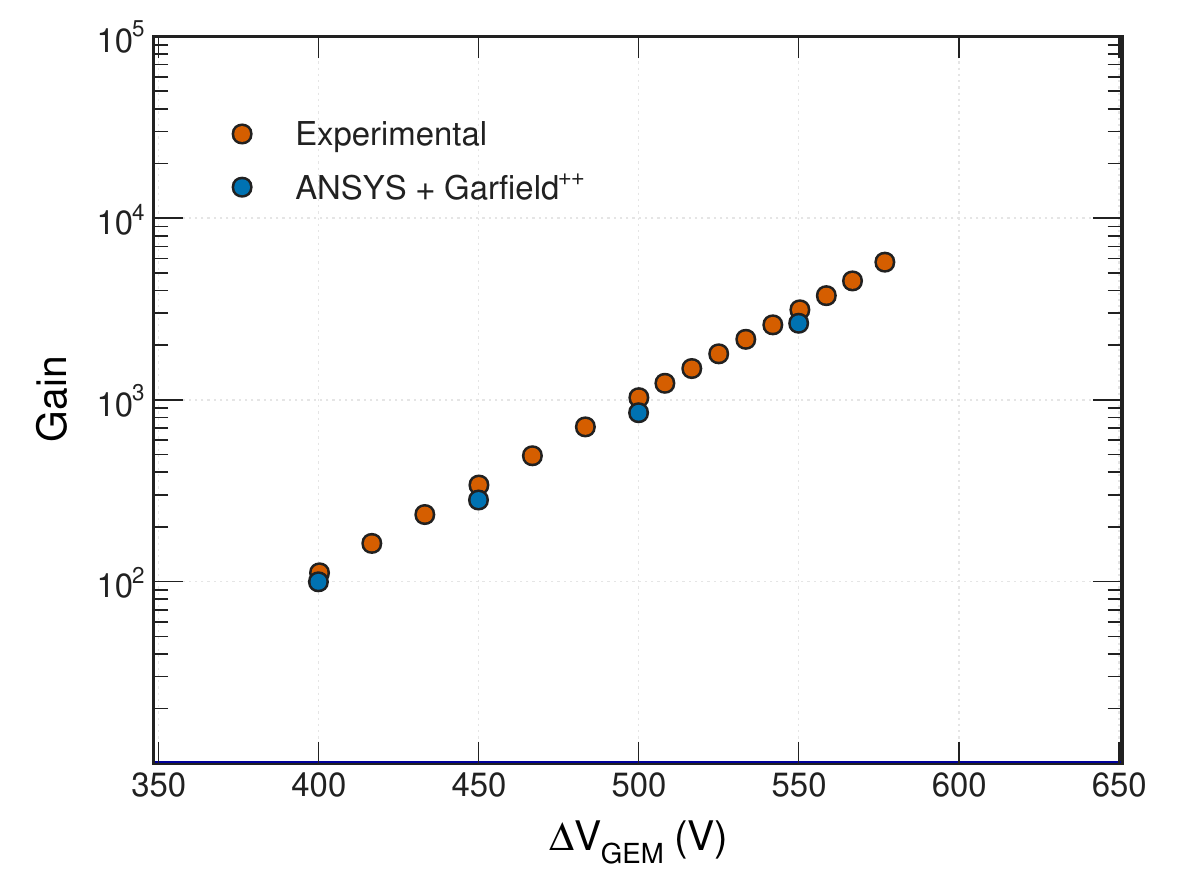}
  \caption{(Color online) Effective gain of a single biconical GEM as a function of
  $\Delta V_{\mathrm{GEM}}$. Orange circles: experimental data from
  Ref.~\cite{bachmann1999charge}; blue circles: ANSYS~+~Garfield$^{++}$ predictions.}
  \label{fig:ExpVsSim}
\end{figure}

The comparison is shown in Figure~\ref{fig:ExpVsSim}. The simulation reproduces the
measured gain to within approximately 20\% over the full voltage range investigated,
confirming that the framework captures the essential physics of electron transport and
avalanche multiplication within GEM holes. This level of agreement is consistent with
similar ANSYS--Garfield$^{++}$ studies in the
literature~\cite{Angiras:2025fcr,mondal2025optimizing} and provides a sound basis for the
parametric investigations that follow.

\subsection{Ion Absorption in the Biconical Geometry}
\label{sec:biconical}

\begin{figure}[!t]
  \centering
  \includegraphics[width=\columnwidth]{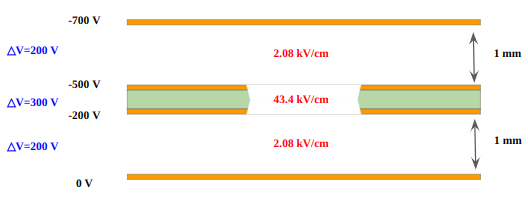}
  \caption{(Color online) Electrostatic analysis of a single-foil biconical GEM in ANSYS
  Maxwell. Inner diameter 50~$\mu$m, outer diameter 70~$\mu$m; equal 1~mm drift and
  induction gaps; $\Delta V = 300$~V across the foil.}
  \label{fig:Biconical_SingleGEM}
\end{figure}

\begin{figure}[!t]
  \centering
  \includegraphics[width=\columnwidth]{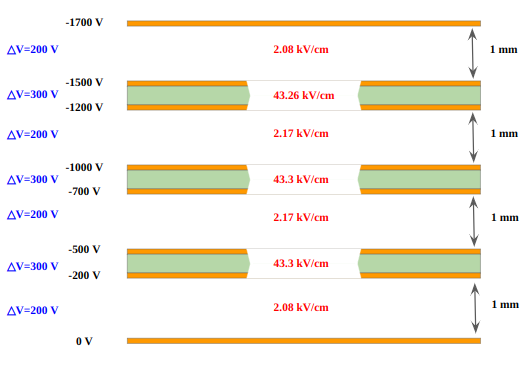}
  \caption{(Color online) Electrostatic analysis of a triple-foil biconical GEM in ANSYS
  Maxwell. The three foils are uniformly spaced; hole geometry and applied voltages are
  identical on every foil.}
  \label{fig:Biconical_TripleGEM}
\end{figure}

The biconical hole is the standard GEM geometry. In the present study, the inner and outer
diameters are 50~$\mu$m and 70~$\mu$m, respectively, on a 50~$\mu$m Kapton substrate clad
with 5~$\mu$m copper on each side, at a pitch of 140~$\mu$m
(Table~\ref{tab:sim_parameters}). The electrostatic layout of the single-foil and triple-foil
configurations are shown in Figures~\ref{fig:Biconical_SingleGEM}
and~\ref{fig:Biconical_TripleGEM}, respectively.

\subsubsection{Layer-Resolved Formalism}
\label{sec:formalism}

To analyse ion absorption systematically, we introduce a two-level description. Consider a
detector with $N$ GEM foils indexed $i = 1, \dots, N$ from the drift side. Ions are produced
predominantly in foil $N$ (the readout-side foil, where the avalanche gain is highest). As
these ions drift upstream, they may be absorbed on foil $N$, on any upstream foil $N-1,
N-2, \dots, 1$, or escape into the drift region. We define two complementary quantities.

The \emph{conditional} absorption probability of foil $i$,
\begin{equation}
a_i = P(\text{absorption at foil } i \mid \text{ion reaches foil } i),
\label{eq:conditional_absorption}
\end{equation}
characterises the local ion-collection efficiency of foil $i$ independently of how many
ions actually reach it.

The \emph{absolute} absorption contribution of foil $i$,
\begin{equation}
A_i = \frac{N_{\text{ions absorbed on the }i\text{-th foil}}}{N_{\text{ions produced in the }N\text{-th foil}}} \times 100\%,
\label{eq:Ai_definition}
\end{equation}
is the fraction of the original ion population (from foil $N$) that terminates on foil $i$.
An ion produced in foil $N$ reaches foil $i$ only if it is not absorbed by any of the foils
$i+1, \dots, N$ it encounters first, so the two quantities are related by
\begin{equation}
A_i = a_i \prod_{j=i+1}^{N} (1 - a_j),
\label{eq:absolute_vs_conditional}
\end{equation}
with the empty product equal to unity for $i = N$. The escape probability the fraction of
ions from foil $N$ that reach the drift region follows the multiplicative law
\begin{equation}
P_{\text{escape}} = \prod_{i=1}^{N} (1 - a_i),
\label{eq:P_escape}
\end{equation}
and the total absorption of last-stage ions is
\begin{equation}
A_{\text{total}} = \sum_{i=1}^{N} A_i = 1 - P_{\text{escape}}.
\label{eq:A_total_last_stage}
\end{equation}
The distinction between $a_i$ and $A_i$ is physically important. In the biconical quintuple
GEM, for example, foil $N-1$ has an absolute contribution of only $A_{N-1} = 3.86\%$, yet
its conditional absorption probability is $a_{N-1} \approx 7.9\%$, because fewer than half
of the ions from foil $N$ actually reach it ($1 - a_N = 48.61\%$).

\subsubsection{Sequential Absorption Results}

\begin{figure}[!t]
  \centering
  \includegraphics[width=\columnwidth]{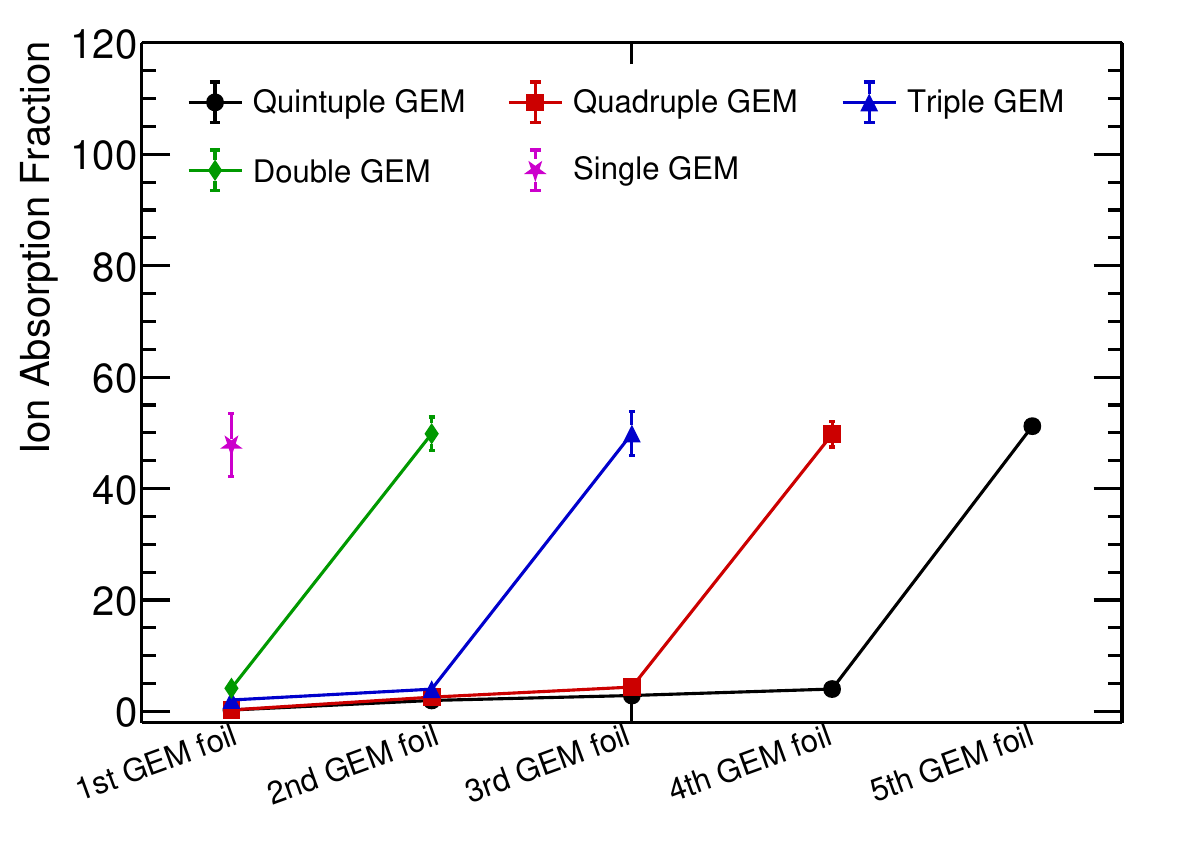}
  \caption{(Color online) Ion absorption contribution $A_i$ of each GEM foil for biconical
  configurations for ions produced in the last foil. Foils are numbered from the drift
  side. Symbols: black circles quintuple GEM; red squares quadruple; blue triangles —
  triple; green diamonds double; magenta star single.}
  \label{fig:biconical_ion_absorption}
\end{figure}

Figure~\ref{fig:biconical_ion_absorption} shows the per-foil absorption contribution $A_i$
for single- through quintuple-GEM detectors with biconical holes, and the corresponding
values are listed in Table~\ref{Table:biconical_ion_absorption_last_stage_production}. The
pattern is consistent across all multiplicities: the production foil (foil $N$) captures
roughly 50\% of the ions it generates, while each successive upstream foil adds a smaller
but non-negligible contribution, ranging from 0.2\% to about 4\%.

\begin{table}[!t]
\centering
\resizebox{\columnwidth}{!}{
\begin{tabular}{|c|c|c|c|c|c|}
\hline
\multirow{2}{*}{\shortstack{Absorbing\\foil}} & \multicolumn{5}{c|}{Detector Configuration} \\
\cline{2-6}
 & \shortstack{Single\\GEM} & \shortstack{Double\\GEM} & \shortstack{Triple\\GEM} & \shortstack{Quadruple\\GEM} & \shortstack{Quintuple\\GEM} \\
\hline
$N$ (production) & 47.87\% & 49.85\% & 49.46\% & 49.68\% & 51.39\% \\
\hline
$N-1$ & - & 3.90\% & 3.83\% & 3.77\% & 3.86\% \\
\hline
$N-2$ & - & - & 2.59\% & 2.58\% & 2.76\% \\
\hline
$N-3$ & - & - & - & 0.24\% & 1.91\% \\
\hline
$N-4$ & - & - & - & - & 0.23\% \\
\hline
\end{tabular}%
}
\caption{Absolute ion absorption contributions $A_i$ (Eq.~\ref{eq:Ai_definition}) for
the biconical geometry. Ions are produced in the readout-side foil $N$ and are absorbed
either on that foil or on the upstream foils $N-1, N-2, \dots$}
\label{Table:biconical_ion_absorption_last_stage_production}
\end{table}

The physical origin of this hierarchy is straightforward. Inside the GEM holes, the electric
field is strongly non-uniform: field lines converge onto the copper electrodes, efficiently
funneling locally produced ions toward the foil surface. In the transfer and drift regions
above the stack, by contrast, the field is nearly uniform and field lines connect successive
foils rather than terminating on electrodes, allowing ions that escape the production foil to
drift upstream with comparatively little additional interception. Adding foils therefore
improves absorption through a cumulative, multiplicative filtering mechanism, as expressed
by Eq.~\eqref{eq:P_escape}: each extra foil reduces the escape probability by a factor
$(1-a_i)$, with diminishing returns as the probability is already small. The total
last-stage absorption $A_{\text{total}}$ rises from $\approx\!48\%$ for a single GEM to
$\approx\!62\%$ for a quintuple GEM (red markers in Figure~\ref{fig:ion_absorption}),
consistent with earlier multi-GEM studies~\cite{bondar2003study,patra2018characteristic,greene2022effective}.

\subsubsection{Detector-Level Absorption Metric}
\label{sec:overall_metric}

The last-stage quantity $A_{\text{total}}$ describes only ions produced in foil $N$. In a
real detector, every GEM foil undergoes gas multiplication and generates its own ion
population; ions produced in upstream foils travel a shorter path to the drift region and
are intrinsically more likely to escape. A metric restricted to the final stage alone
therefore underestimates the true ion load. We define the \emph{overall} absorption
fraction
\begin{equation}
A_{\text{overall}} = \frac{\text{Total ions absorbed in the detector}}{\text{Total ions generated in all GEM foils}},
\label{eq:overall_absorption_definition}
\end{equation}
which counts all ions regardless of their stage of origin and is the physically relevant
quantity governing space-charge buildup in the drift volume. For a detector with $n$ foils,
letting $N_i$ denote the ion yield of foil $i$ and $A_{i\rightarrow j}$ the fraction of
ions from foil $i$ absorbed on foil $j \leq i$, this becomes
\begin{equation}
A_{\text{overall}} = \frac{\displaystyle\sum_{i=1}^{n} N_i \left( \sum_{j=1}^{i} A_{i\rightarrow j} \right)}{\displaystyle\sum_{i=1}^{n} N_i}.
\label{eq:overall_general_formula}
\end{equation}
Note that $\sum_{j} A_{i\rightarrow j}$ is precisely $A_{\text{total}}$ of
Eq.~\eqref{eq:A_total_last_stage} applied to ions originating in foil $i$.

To make this concrete, consider the double-GEM case. Let $N_1$ and $N_2$ denote the ion
yields of the upper and lower foils, respectively. Ions from foil~2 are absorbed
at 49.84\% by foil~2 itself and a further 4.13\% by foil~1:
\begin{equation}
N^{\text{abs}}_2 = (0.4984 + 0.0413)\,N_2 = 0.5397\,N_2,
\label{eq:A2}
\end{equation}
with the remainder escaping toward the drift region:
\begin{equation}
N^{\text{esc}}_2 = N_2 - N^{\text{abs}}_2 = 0.4603\,N_2.
\label{eq:E2}
\end{equation}
Ions from foil~1, which face only a single barrier before reaching the drift region, are
absorbed at 47.83\%:
\begin{equation}
N^{\text{abs}}_1 = 0.4783\,N_1, \qquad N^{\text{esc}}_1 = 0.5217\,N_1.
\label{eq:A1E1}
\end{equation}
The detector-level absorption fraction is therefore
\begin{equation}
A_{\text{overall}} = \frac{0.4783\,N_1 + 0.5397\,N_2}{N_1 + N_2}.
\label{eq:overall_double_gem}
\end{equation}
under cascaded operation, $N_2 \gg N_1$ because the gain of the second stage amplifying
the electron cloud already multiplied by the first greatly exceeds that of the first foil:
\begin{equation}
N_2 \approx G_2 \, T_{1\rightarrow 2} \, N_{e,1},
\label{eq:N2_gain_relation}
\end{equation}
where $G_2$ is the gain of the second stage, $T_{1\rightarrow 2}$ the interstage electron
transfer efficiency, and $N_{e,1}$ the electron population entering the second foil. As a
result, $A_{\text{overall}}$ lies close to, but slightly below, the last-stage value, pulled
downward by the smaller, lower-probability contribution from foil~1. This pattern generalises
to any number of foils: the detector-level performance is dominated by the final amplification
stages, while upstream contributions reduce $A_{\text{overall}}$ by a small but
physically meaningful amount.

The overall ion-absorption fractions for the biconical geometry are 47.83\%, 52.57\%,
54.70\%, 56.12\%, and 59.06\% for single- through quintuple-GEM detectors (red points in
Figure~\ref{fig:overall_ion_absorption}). The monotonic increase with multiplicity follows
directly from the multiplicative escape-probability law: each additional foil can only
reduce or leave unchanged the net escape probability. The diminishing increments at higher
multiplicities an unavoidable consequence of operating on an already-small escape
probability indicate that cascading alone is an increasingly inefficient route to further
IBF reduction once three or four foils are in place.

\subsection{Ion Absorption in the Single-Conical Geometry}
\label{sec:single_conical}

\begin{figure}[!t]
  \centering
  \includegraphics[width=\columnwidth]{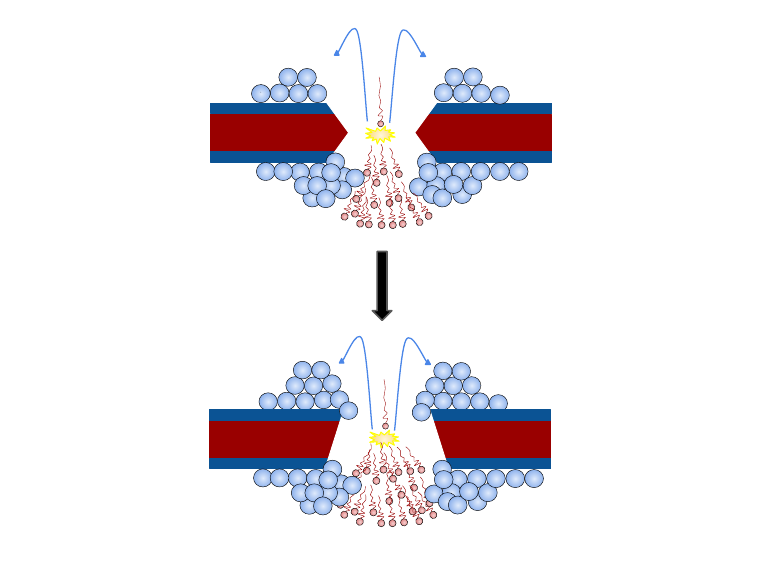}
  \caption{(Color online) Schematic comparison of ion-collection mechanisms. Top:
  biconical geometry the multiplication zone is far from the upper electrode, and ions
  (blue spheres) can escape into the drift region (blue arrows) before being collected.
  Bottom: single-conical geometry. The asymmetric hole profile brings the upper electrode
  closer to the avalanche, collecting most ions before they can drift upward. Red lines
  represent electron pathways.}
  \label{fig:modified_geo}
\end{figure}

The single-conical hole profile addresses a geometric limitation of the biconical design.
In a biconical GEM, the waist of the hourglass-shaped hole is located near the centre of
the polyimide substrate, placing the multiplication zone at some distance from both copper
electrodes. Positive ions generated in the avalanche therefore travel a comparatively long
path before reaching an electrode, increasing their probability of escaping into the
transfer region. In the single-conical geometry, the hole tapers from a smaller diameter
at the top (50~$\mu$m) to a larger diameter at the bottom (70~$\mu$m), shifting the
avalanche region closer to the upper copper electrode. The asymmetric field distribution
within the hole guides ions toward the nearby electrode surface almost immediately after
their creation, substantially shortening the collection path. A schematic comparison of
the two ion-collection mechanisms is shown in Figure~\ref{fig:modified_geo}.

\begin{figure}[!t]
  \centering
  \includegraphics[width=\columnwidth]{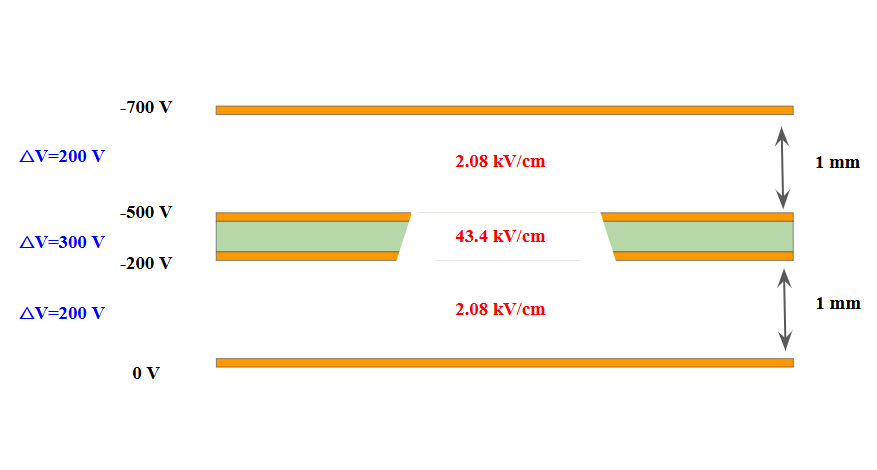}
  \caption{(Color online) Electrostatic analysis of a single-foil single-conical GEM in
  ANSYS Maxwell. Upper diameter 50~$\mu$m, lower diameter 70~$\mu$m; equal 1~mm drift and
  induction gaps; $\Delta V = 300$~V.}
  \label{fig:SingleConical_SingleGEM}
\end{figure}

In addition to improved ion collection, the single-conical design eliminates the
intermediate constriction of biconical holes, providing a less obstructed path for
avalanche electrons and yielding a measurable enhancement in effective
gain~\cite{Angiras:2025fcr,mondal2025optimizing,rana2026designoptimizationtriplegas,Sarkar:2025tvu}. The single-conical geometry also offers
a practical fabrication advantage: single-sided etching with a single photolithographic
mask removes the alignment requirement of double-mask biconical processes, reducing
the risk of hole asymmetry due to mask
misregistration~\cite{villa2011progress}.

\subsubsection{Sequential Absorption Results}

\begin{figure}[!t]
  \centering
  \includegraphics[width=\columnwidth]{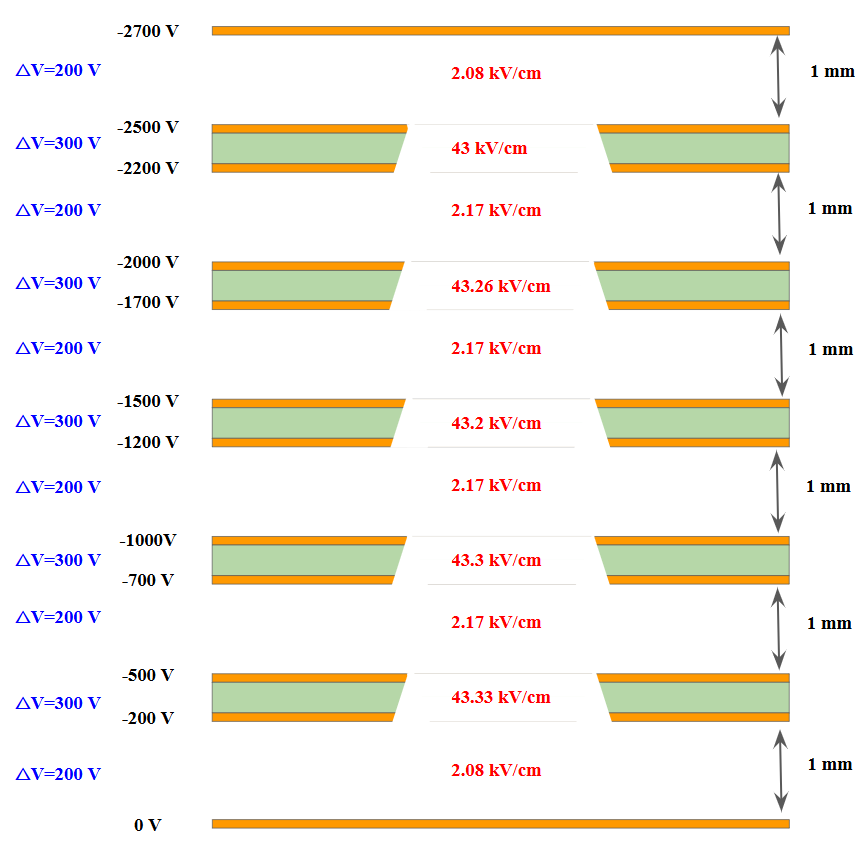}
  \caption{(Color online) Electrostatic analysis of a quintuple-foil single-conical GEM
  in ANSYS Maxwell. Five equally spaced foils (1~mm gaps); identical hole geometry and
  voltages on every foil.}
  \label{fig:SingleConical_PentaGEM}
\end{figure}

\begin{figure}[!t]
  \centering
  \includegraphics[width=\columnwidth]{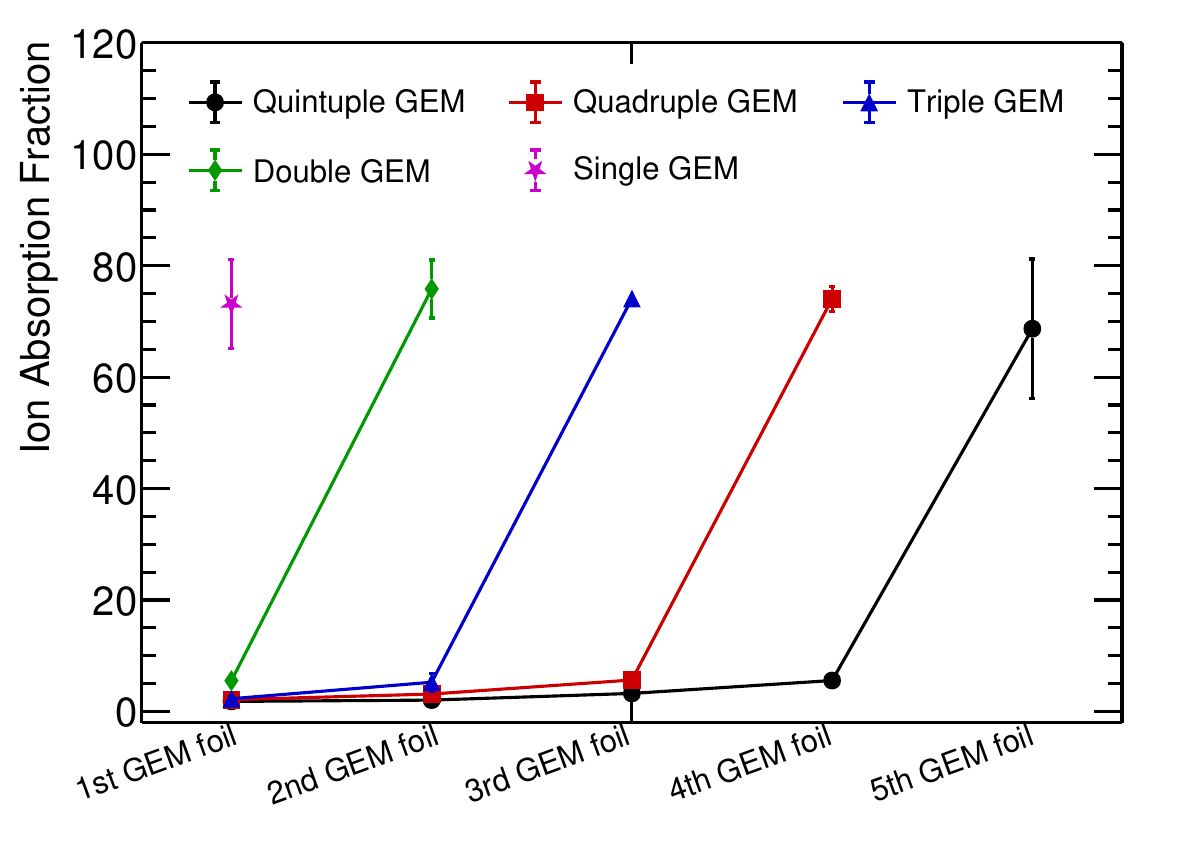}
  \caption{(Color online) Ion absorption contribution $A_i$ of each GEM foil for
  single-conical configurations, for ions produced in the last foil. Symbol convention
  as in Figure~\ref{fig:biconical_ion_absorption}.}
  \label{fig:singleconical_ion_absorption}
\end{figure}

The electrostatic geometry for a single-conical quintuple-GEM is shown in
Figure~\ref{fig:SingleConical_PentaGEM}. The per-foil absorption contributions for
single-conical detectors are presented in Figure~\ref{fig:singleconical_ion_absorption}
and Table~\ref{Table:singleconical_ion_absorption_last_stage_production}.

\begin{table}[!t]
\centering
\resizebox{\columnwidth}{!}{
\begin{tabular}{|c|c|c|c|c|c|}
\hline
\multirow{2}{*}{\shortstack{Absorbing\\foil}} & \multicolumn{5}{c|}{Detector Configuration} \\
\cline{2-6}
 & \shortstack{Single\\GEM} & \shortstack{Double\\GEM} & \shortstack{Triple\\GEM} & \shortstack{Quadruple\\GEM} & \shortstack{Quintuple\\GEM} \\
\hline
$N$ (production) & 74.93\% & 75.05\% & 74.85\% & 74.08\% & 74.17\% \\
\hline
$N-1$ & - & 5.27\% & 5.14\% & 5.54\% & 5.60\% \\
\hline
$N-2$ & - & - & 2.91\% & 3.55\% & 2.86\% \\
\hline
$N-3$ & - & - & - & 2.31\% & 2.24\% \\
\hline
$N-4$ & - & - & - & - & 1.88\% \\
\hline
\end{tabular}%
}
\caption{Absolute ion absorption contributions $A_i$ (Eq.~\ref{eq:Ai_definition}) for
the single-conical geometry. Convention as in
Table~\ref{Table:biconical_ion_absorption_last_stage_production}.}
\label{Table:singleconical_ion_absorption_last_stage_production}
\end{table}

\begin{figure}[!t]
  \centering
  \includegraphics[width=\columnwidth]{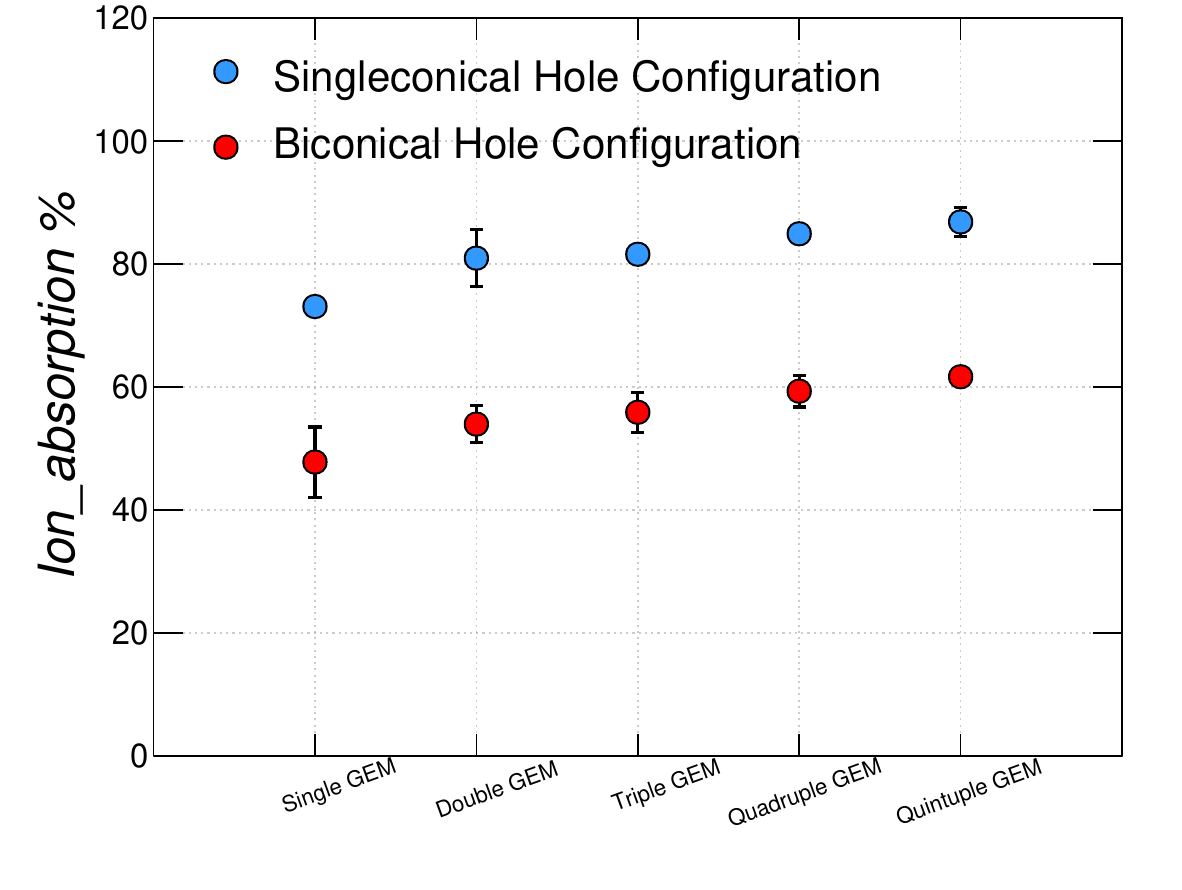}
  \caption{(Color online) Total last-stage ion absorption $A_{\text{total}}$
  (y-axis: ion absorption fraction, \%) for single- through quintuple-GEM detectors.
  Red markers: biconical geometry; blue markers: single-conical geometry. Both geometries
  show monotonically increasing absorption with multiplicity, with diminishing increments
  at higher foil numbers.}
  \label{fig:ion_absorption}
\end{figure}

The most striking feature is the absorption fraction at the production foil: approximately
75\% for single-conical holes, compared with roughly 50\% for biconical holes. This
$\approx\!25$\%-point improvement reflects the shorter ion-collection path in the
asymmetric field geometry. Upstream foils continue to filter ions cumulatively contributing
between $\sim\!2\%$ and $\sim\!6\%$ per foil confirming that the multi-stage suppression
mechanism remains active and is, in fact, strengthened by the geometry change. The total
last-stage absorption $A_{\text{total}}$ for the single-conical geometry (blue markers in
Figure~\ref{fig:ion_absorption}) exceeds that of the biconical geometry by 25--27\% at every multiplicity.

\subsubsection{Detector-Level Absorption}

\begin{figure}[!t]
  \centering
  \includegraphics[width=\columnwidth]{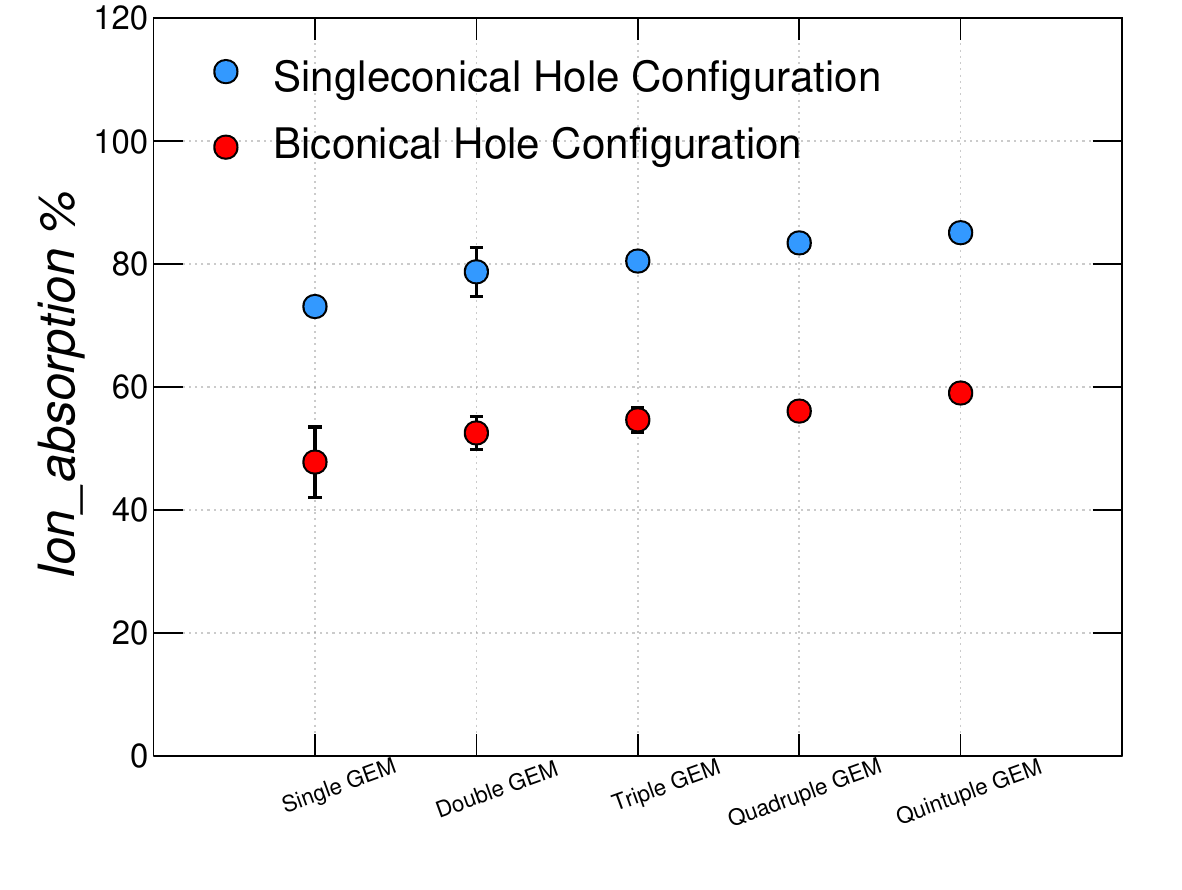}
  \caption{(Color online) Overall ion-absorption fraction $A_{\text{overall}}$
  (y-axis: ion absorption fraction, \%) accounting for ions generated throughout the entire
  GEM stack for single- through quintuple-GEM detectors. Red markers: biconical geometry;
  blue markers: single-conical geometry. Both geometries show monotonically increasing
  absorption with multiplicity; the single-conical geometry is consistently
  $\approx\!25$\% points higher.}
  \label{fig:overall_ion_absorption}
\end{figure}

Applying the formalism of Section~\ref{sec:overall_metric} to the single-conical geometry,
the overall ion-absorption fraction $A_{\text{overall}}$ increases from 72.19\% for a
single-GEM detector to $\approx$78.8\%, 80.5\%, 83.4\%, and 85.1\% for the double-,
triple-, quadruple-, and quintuple-GEM configurations, respectively (blue points in
Figure~\ref{fig:overall_ion_absorption}). These values exceed those of the biconical geometry
by approximately 25\% points at every multiplicity the same improvement seen at
the single-foil level, confirming that the gain in local ion-collection efficiency translates
directly into detector-level performance. Notably, although the higher gain of the
single-conical geometry produces more ions in absolute terms, the substantially higher
absorption probability more than compensates: the net ion load reaching the drift region is
reduced, not increased.

A comparison of the two geometries across all configurations leads to three key quantitative
conclusions. First, the absorption at the production foil increases from $\approx\!50\%$ to
$\approx\!70$--75\% upon switching to the single-conical profile. Second, the total
last-stage absorption $A_{\text{total}}$ rises by 25-27\% for every
multiplicity. Third, the detector-level improvement is approximately 25\% points
even after accounting for the higher ion production associated with the elevated gain.
Taken together, the improvement achieved by changing the hole geometry in a single-GEM
detector exceeds what can be obtained by adding four additional biconical foils making
the hole profile the more effective design lever for ion suppression in GEM-based detectors.

\subsection{IBF/Gain Ratio and GEM Multiplicity}
\label{sec:IBFbyGain}

A complete characterisation of GEM performance must distinguish ion suppression from gain.
A configuration with a lower total gain necessarily produces fewer ions in absolute terms,
regardless of its actual suppression efficiency. To remove this trivial dependence, we use
the ratio IBF/Gain, the number of back-drifting ions per electron collected at the readout as the primary figure of merit~\cite{ball2014ion,greene2022effective}. Throughout this
work, IBF denotes the number of avalanche ions reaching the drift cathode and $G$ the
effective gain, both determined for the same primary-electron ensemble.

\begin{figure}[!t]
  \centering
  \includegraphics[width=\columnwidth]{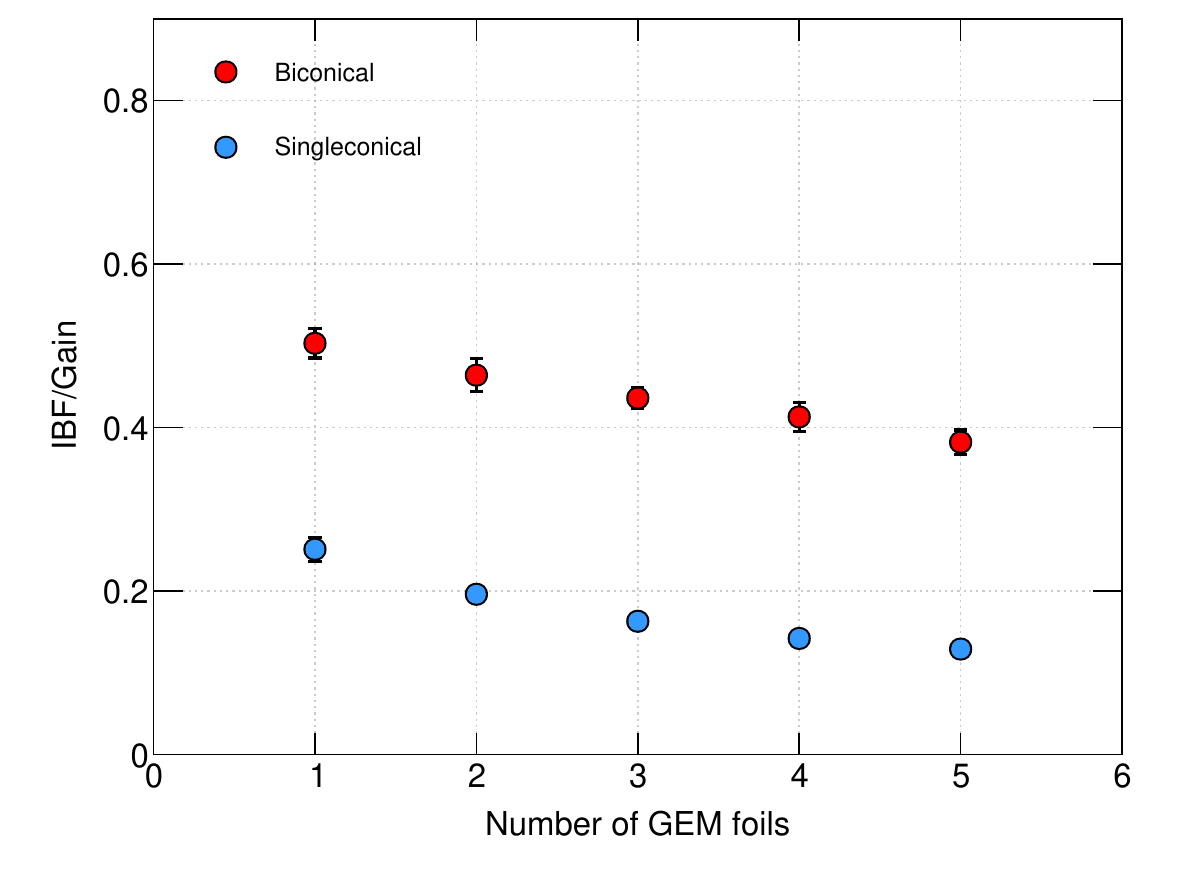}
  \caption{(Color online) IBF/Gain ratio as a function of the number of GEM foils.
  Red circles: biconical; blue circles: single-conical. The single-conical geometry
  achieves a factor-of-two to three reduction at every multiplicity.}
  \label{fig:IBFbyGain}
\end{figure}

The IBF/Gain ratio as a function of GEM multiplicity is shown in
Figure~\ref{fig:IBFbyGain}. For the biconical geometry, the ratio decreases approximately
linearly from $\approx\!0.50$ (single GEM) to $\approx\!0.38$ (quintuple GEM). For the
single-conical geometry, the ratio is lower by a factor of two to three at every
multiplicity, falling from $\approx\!0.25$ to $\approx\!0.13$; the dependence is nonlinear
and approaches saturation, with the largest reduction occurring between one and two foils
and only a marginal change between four and five. Since the ratio is normalised to the gain,
this improvement cannot be attributed to a reduced avalanche size: it reflects a genuine
enhancement in ion suppression relative to the electron signal. The saturation behaviour of
the single-conical curve further suggests that the hole geometry rather than additional
foils is the primary driver of IBF/Gain reduction.

\section{Extension to a Realistic Triple-GEM: Effect of the Second Transfer Gap}
\label{sec:transfer_gap}

The preceding analysis was carried out under a deliberately idealised, uniform gap
configuration (all gaps equal to 1~mm) to isolate the effects of multiplicity and hole
geometry. Real triple-GEM detectors typically operate with a non-uniform gap hierarchy
commonly in a 3:1:2:1 ratio for the drift, first transfer, second transfer, and induction
regions~\cite{CMS:2015GEMTDR} in which the transfer gaps affect both electron
extraction efficiency and the ion drift path between
foils~\cite{bachmann1999charge,bhattacharya2017simulation,natal2018ion}. To test whether
the advantage of the single-conical geometry survives under these more realistic conditions,
we vary the
second transfer-gap length between 1.2 and 3.5~mm for both hole geometries and evaluate
the gain, IBF, and IBF/Gain ratio for each configuration.

\subsection{Conventional Biconical GEM}

\begin{figure}[!t]
  \centering
  \includegraphics[width=\columnwidth]{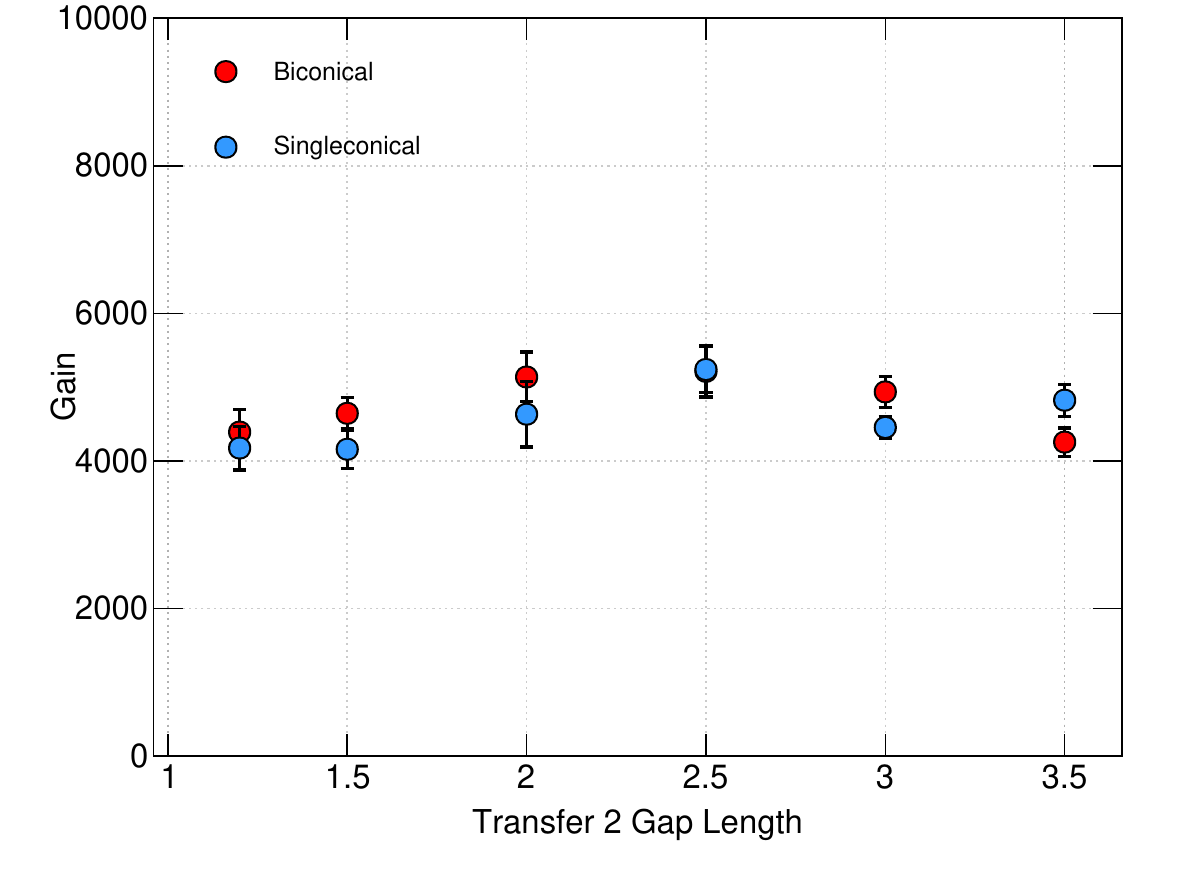}
  \caption{(Color online) Effective gain versus second transfer-gap length for the
  biconical (red) and single-conical (blue) geometries. Both show a similar non-monotonic
  trend with a maximum near 2.0--2.5~mm, indicating that gain is governed by the
  transfer-gap geometry rather than the hole profile. Error bars represent statistical
  uncertainty.}
  \label{fig:GainVsT2}
\end{figure}

\textbf{Effective Gain.}
The effective gain of the biconical triple-GEM shows a non-monotonic dependence on the
second transfer gap (Figure~\ref{fig:GainVsT2}). The gain rises from $\approx\!4400$ at a
gap of 1.2~mm to a maximum of $\approx\!5200$ near 2.0--2.5~mm, then decreases gradually
at larger gaps. The initial rise reflects improved electron extraction and focusing as the
gap widens beyond the region where field lines from GEM2 terminate prematurely on the upper
copper electrode of GEM3. The subsequent decline arises from transverse diffusion: at gaps
exceeding $\approx\!2.5$~mm, the electron cloud broadens sufficiently during transit that
an increasing fraction of electrons misses the GEM3 holes and is collected on the copper
surface rather than entering the avalanche region.

\begin{figure}[!t]
  \centering
  \includegraphics[width=\columnwidth]{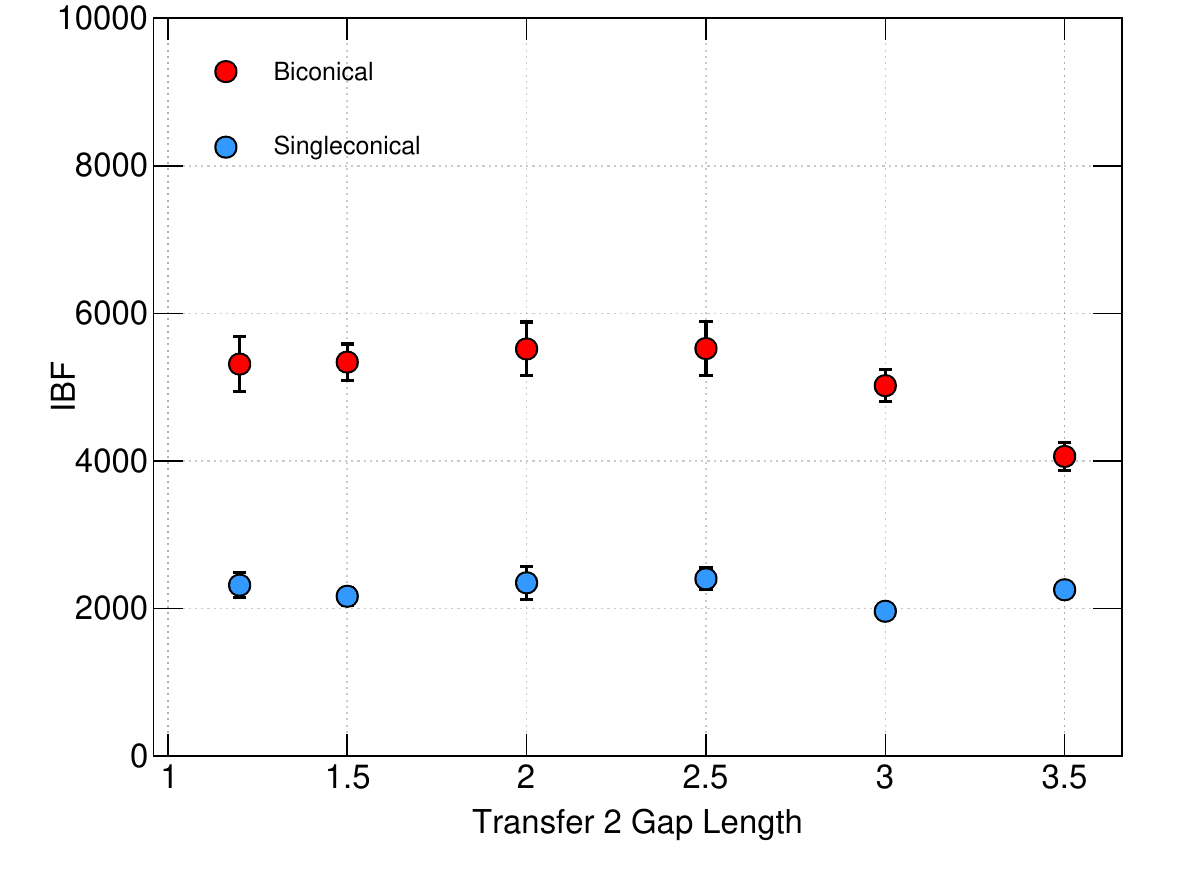}
  \caption{(Color online) Ion backflow (IBF) versus second transfer-gap length. The
  biconical configuration shows consistently higher IBF; the single-conical geometry
  remains substantially lower across the full range, reflecting its intrinsic ion-blocking
  capability. Error bars represent statistical uncertainty.}
  \label{fig:IBFVsT2}
\end{figure}

\textbf{Ion Backflow.}
The absolute IBF follows the gain qualitatively, rising modestly up to 2.0--2.5~mm before
declining at larger gap lengths (Figure~\ref{fig:IBFVsT2}). Since the number of ions
produced in an avalanche scales directly with the avalanche size, the IBF peak coincides
with the gain maximum.

\textbf{IBF/Gain Ratio.}
In contrast to the non-monotonic behaviour of gain and absolute IBF, the IBF/Gain ratio
decreases monotonically from $\approx\!1.2$ at 1.2~mm to $\approx\!0.95$ at 3.5~mm
(Figure~\ref{fig:RatioVsT2}). This monotonic decrease is physically significant: at the
largest gap lengths, both gain and absolute IBF decline, but IBF falls by a larger relative
amount than the electron multiplication. A longer transfer gap increases the probability
that ions originating in GEM3 terminate on the lower electrode of GEM2 rather than
continuing to drift toward the cathode; the associated field-geometry modification further
suppresses inter-stage ion transmission. Consequently, despite the modest gain reduction at
large gaps, the ion-blocking efficiency of the biconical detector continues to improve.
Transfer gaps of 3.0--3.5~mm yield the lowest IBF/Gain ratio while still preserving
comparatively high gain.

\subsection{Modified Single-Conical GEM}

\textbf{Detector Gain.}
The gain of the single-conical triple-GEM mirrors that of the biconical geometry closely,
reaching a maximum near 2.5~mm before declining (Figure~\ref{fig:GainVsT2}). This
similarity confirms that modifying the hole profile does not significantly alter the
electron multiplication mechanism; the single-conical design preserves efficient electron
amplification while independently modifying ion transport.

\textbf{Ion Backflow.}
A pronounced difference between the two geometries emerges in the IBF behaviour. Across
the full range of transfer-gap lengths, the single-conical detector produces approximately
45--60\% less absolute IBF than the biconical configuration: while the biconical geometry
yields $\approx\!4000$--5500 ions, the single-conical geometry produces only
$\approx\!2000$--2400 (Figure~\ref{fig:IBFVsT2}). Since the gains are comparable at every
gap length (Figure~\ref{fig:GainVsT2}), this reduction is not a consequence of a smaller
avalanche. It originates instead from the asymmetric field distribution inside the
single-conical hole: the avalanche develops closer to the upper copper electrode, so ions
are collected almost immediately after their creation before they can drift into the
transfer region.

\begin{figure}[!t]
  \centering
  \includegraphics[width=\columnwidth]{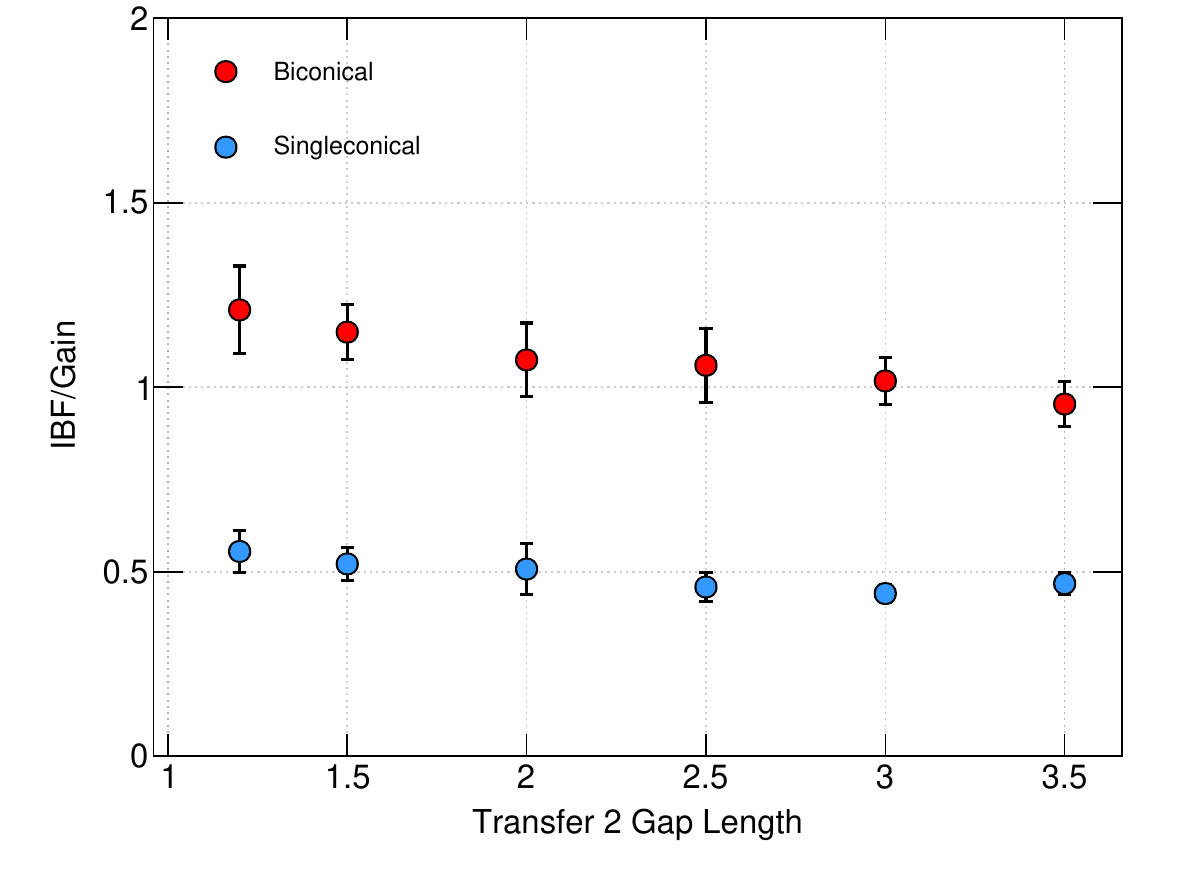}
  \caption{(Color online) IBF/Gain ratio versus second transfer-gap length. The
  single-conical geometry maintains a factor-of-two lower ratio throughout, confirming
  that the ion-suppression advantage originates from the hole geometry rather than the
  gap configuration. Error bars represent statistical uncertainty.}
  \label{fig:RatioVsT2}
\end{figure}

\textbf{IBF/Gain Ratio.}
The IBF/Gain ratio of the single-conical detector decreases from $\approx\!0.55$ to
$\approx\!0.45$ across the investigated gap range, compared with 0.95--1.2 for the
biconical configuration, a factor-of-two reduction that is maintained at every gap length
(Figure~\ref{fig:RatioVsT2}). Because this improvement is achieved without any penalty in
gain, it reflects a genuine enhancement in ion suppression relative to the useful electron
signal. The ratio reaches its minimum near a transfer gap of $\approx\!3$~mm, suggesting
that larger gaps further enhance the intrinsic ion-trapping capability of the asymmetric
hole geometry. The central finding of the main study that the single-conical profile
substantially outperforms the biconical design therefore persists in full under a more
realistic, non-uniform gap configuration.

\section{Discussion}
\label{sec:discussion}

The results presented in Sections~\ref{sec:biconical}--\ref{sec:transfer_gap} carry
several implications for the design and operation of GEM-based detectors, which we
discuss here in the context of practical applications.

\subsection{Design Implications for High-Rate Detectors}

The comparison of the biconical and single-conical geometries reveals a clear hierarchy:
the hole geometry, not the number of foils, is the dominant lever for controlling ion
backflow. A single-conical single-GEM achieves an overall ion-absorption fraction
($A_{\text{overall}} = 72.19\%$) that already exceeds that of a biconical quintuple-GEM
(59.06\%), while simultaneously delivering a factor of two reduction in the IBF/Gain
ratio. For applications such as the ALICE TPC~\cite{ALICE:2013TPCTDR,ALICETPC:2021upgrade},
where the IBF/Gain ratio must be kept below $\sim\!10^{-3}$ to control space-charge
distortions at LHC collision rates, these results suggest that optimising the hole
geometry could reduce the required number of foils while maintaining or improving ion
suppression with direct benefits for detector complexity, material budget, and cost.

\subsection{Relevance to the Transfer-Gap Optimization}

The transfer-gap study (Section~\ref{sec:transfer_gap}) demonstrates that the single-conical
advantage is not a delicate property of the idealized uniform-gap configuration: it persists
across the full range of second transfer-gap lengths from 1.2 to 3.5~mm, spanning both
the gain-maximum regime and the diffusion-dominated decline. The IBF/Gain ratio of the
single-conical geometry is approximately a factor of two lower than that of the biconical
geometry at every gap length investigated. Furthermore, both geometries favour larger
transfer gaps (3.0--3.5~mm) for minimising IBF/Gain while preserving high gain, a
result broadly consistent with earlier studies~\cite{bhattacharya2017simulation,natal2018ion}
and relevant to the non-uniform 3:1:2:1 gap ratios used in real
detectors~\cite{CMS:2015GEMTDR}.

\subsection{Saturation and the Limit of Cascading}

The diminishing returns of cascading, already visible for the biconical geometry beyond
three foils, are a direct consequence of the multiplicative escape-probability law
(Eq.~\eqref{eq:P_escape}): each additional foil reduces the already-small escape
probability by a further factor $(1-a_i)$, so the marginal benefit decreases with every
stage added. For the single-conical geometry, this saturation is even more pronounced
(Figure~\ref{fig:IBFbyGain}), since the initial absorption at the production foil ($\approx$75\%)
is already high. These observations suggest that beyond three or four single-conical foils,
further gain in ion suppression is better sought through optimisation of the field ratios
and gap geometry rather than through additional foils, a conclusion with practical
significance for detector cost and material budget.

\subsection{Outlook}

The present work establishes a quantitative simulation framework and provides predictions
for ion-absorption fractions and IBF/Gain ratios that can be tested experimentally. Direct
IBF measurements on single-conical GEM foils, which, to our knowledge, have not yet been
reported, would be the most valuable experimental validation. Such measurements could be
performed with standard X-ray sources and current-mode readout~\cite{bachmann1999charge},
adapted to the single-conical foil geometry. Extension of the simulation to include dielectric
charging-up~\cite{alfonsi2012simulation,correia2014dynamic}, space-charge feedback at high
rates, and Ne-based gas mixtures used in large TPC upgrades~\cite{ALICE:2013TPCTDR} are the
natural next steps toward a full predictive model for high-rate detector performance.

\section{Limitations and Scope of the Present Study}
\label{sec:limitations}

The results reported here were obtained under a deliberately controlled set of conditions
chosen to isolate the effects of GEM multiplicity and hole geometry. The following
limitations define the range of validity of the conclusions.

\begin{enumerate}

\item \textbf{Ion transport not independently validated.} The ANSYS--Garfield$^{++}$
framework was benchmarked only against the gain measurements of
Ref.~\cite{bachmann1999charge}, which test electron transport and avalanche multiplication
but not ion transport. No experimental measurements of IBF or ion-absorption fractions
exist for the single-conical geometry, nor for the multi-foil biconical configurations
studied here. The absorption fractions and IBF/Gain ratios reported are predictions
awaiting experimental confirmation.

\item \textbf{Uncharged dielectric.} All field maps correspond to an initially uncharged
polyimide surface. In an operating GEM, charge deposited on the hole walls shifts the
intra-hole field and drives a transient evolution of the gain toward an
equilibrium~\cite{alfonsi2012simulation,correia2014dynamic}. Because the exposed dielectric
surface area differs between biconical and single-conical profiles, the two geometries may
respond differently to charging-up, and the equilibrium operating point may differ from the
initial state studied here.

\item \textbf{Static field; no space-charge feedback.} Each simulated avalanche is
tracked in the electrostatic field of the Laplace solution; the field generated by the
avalanche charge itself and by accumulated back-drifting ions is
neglected~\cite{boehmer2013simulation,ALICE:2013TPCTDR}. The results therefore represent
the low-rate, low-occupancy limit, and any rate dependence of ion absorption is outside the
scope of this study.

\item \textbf{Single gas mixture.} Only Ar/CO$_2$ (70:30) was investigated. Townsend
coefficients, electron diffusion, and ion mobility all depend on the gas composition, and
the absolute values of gain and IBF cannot be transferred directly to other mixtures such
as the Ne-based gases used in large-volume
TPCs~\cite{ALICE:2013TPCTDR}. Whether the single-conical advantage retains its magnitude
in other gases has not been examined.

\item \textbf{Single primary-energy condition.} Simulations used a fixed initial electron
energy of 27~eV throughout. The sensitivity of the results to the primary-electron energy
and to varying distributions of primary ionization (e.g., minimum-ionising versus heavily
ionising particles) was not investigated.

\item \textbf{Fixed operating voltage.} The multiplicity study was conducted at a single
operating point ($\Delta V_{\mathrm{GEM}} = 300$~V on every foil). In practice, detectors
are often operated at higher and unequal foil voltages with field ratios tuned to minimise
IBF~\cite{ball2014ion,ALICE:2013TPCTDR}. The reported values are specific to this operating
point, and IBF-optimised configurations were not explored.

\item \textbf{Idealised uniform gap geometry.} The main study uses equal 1~mm gaps
throughout. While this ensures an unbiased comparison across multiplicities, it does not
correspond to a practical detector~\cite{CMS:2015GEMTDR}. Section~\ref{sec:transfer_gap}
extends the analysis to a varying second transfer gap in the triple-GEM configuration, but
the effects of the drift-gap and induction-gap lengths, and of unequal field ratios, were
not examined.

\end{enumerate}

Despite these constraints, the comparison between the two hole geometries is made under
identical conditions in every case, so these limitations affect both geometries equally
(with the possible exception of charging-up dynamics). The principal conclusion that the
single-conical profile collects a substantially larger fraction of avalanche ions than the
biconical profile is a relative statement and is therefore expected to be robust.
Extending the study to include charging-up effects, space-charge feedback, additional gas
mixtures, and optimised voltage and gap configurations, together with experimental IBF
measurements on single-conical foils, are natural and important next steps.

\section{Summary and Conclusions}
\label{sec:summary}

We have presented a systematic ANSYS--Garfield$^{++}$ simulation study of ion absorption
and backflow in cascaded GEM detectors, covering single- to quintuple-foil configurations
with both the conventional biconical and the modified single-conical hole geometry. The
main conclusions are as follows.

\begin{enumerate}

\item \textbf{Framework validation.} The simulated effective gain of a single biconical GEM
agrees with the experimental measurements of Ref.~\cite{bachmann1999charge} to within
$\approx\!20\%$ over the full $\Delta V_{\mathrm{GEM}}$ range studied, supporting the
reliability of the framework for the parametric investigations reported here.

\item \textbf{Multiplicity dependence.} For the biconical geometry, increasing the number
of foils raises the overall ion-absorption fraction from 47.83\% (single GEM) to 59.06\%
(quintuple GEM), while the IBF/Gain ratio decreases approximately linearly from $\approx\!0.50$
to $\approx\!0.38$. The improvement diminishes with each additional foil, as expected from
the multiplicative escape-probability law.

\item \textbf{Single-conical versus biconical geometry.} Under identical operating
conditions, the single-conical geometry raises the overall ion-absorption fraction by
$\approx\!25$\% points at every multiplicity, reaching 72.19\%--85\% for
single- through quintuple-GEM detectors, and reduces the IBF/Gain ratio by a factor of two
to three. This improvement achieved by changing the hole profile alone exceeds what can
be obtained by adding four further biconical foils. The ion-suppression capability of a
cascaded GEM detector is therefore governed more strongly by the local ion-collection
properties of each foil than by the number of foils.

\item \textbf{Layer-resolved absorption structure.} The foil in which ions are produced
dominates their collection: $\approx\!50\%$ for biconical and $\approx\!70$--75\% for
single-conical holes. Upstream foils add smaller cumulative contributions through the
multiplicative filtering mechanism of Eq.~\eqref{eq:P_escape}. The detector-level
absorption (Eq.~\eqref{eq:overall_general_formula}) lies slightly below the last-stage
value and is governed primarily by the final amplification stages, where ion production is
highest. For the single-conical geometry, the enhancement in absorption more than
compensates for the additional ion production associated with the higher gain.

\item \textbf{Transfer-gap dependence.} In a triple-GEM detector, the effective gain
exhibits a non-monotonic dependence on the second transfer-gap length, peaking near
2.0--2.5~mm, while the IBF/Gain ratio decreases monotonically from 1.2 to 0.95 (biconical)
and from 0.55 to 0.45 (single-conical) across the 1.2--3.5~mm range. The single-conical
geometry reduces the absolute IBF by 45--60\% at comparable gain, maintaining a
factor-of-two reduction in the IBF/Gain ratio at every gap length.

\end{enumerate}

The single-conical hole geometry can also be produced with a simpler single-mask etching
process, removing the alignment requirement of conventional biconical fabrication, although
an experimental demonstration of this advantage and a direct measurement of IBF for
single-conical foils remain to be performed. Subject to the limitations discussed in
Section~\ref{sec:limitations}, these results identify the single-conical geometry as a
promising direction for high-rate gaseous-detector applications in which minimising
space-charge distortions is essential.

\section*{Acknowledgements}
The authors gratefully acknowledge the members of the HEP group at IIT Mandi for stimulating
discussions throughout this work. Financial support from the Anusandhan National Research
Foundation (ANRF), Government of India, through research project ANRF/IRG/2025/000338/PS is
gratefully acknowledged. The authors also thank the Indian Institute of Technology Mandi for
access to its supercomputing facilities, which were essential for completing this study.

\bibliographystyle{elsarticle-harv}
\bibliography{example}

@article{charpak1970some,
  title={Some developments in the operation of multiwire proportional chambers},
  author={Charpak, Georges and Rahm, D and Steiner, H},
  journal={Nuclear Instruments and Methods},
  volume={80},
  number={1},
  pages={13--34},
  year={1970},
  publisher={Elsevier}
}

@article{sauli1977principles,
  title={Principles of operation of multiwire proportional and drift chambers},
  author={Sauli, Fabio},
  journal={Geneva, Switzerland},
  year={1977}
}

@article{charpak1979multiwire,
  title={Multiwire proportional chambers and drift chambers},
  author={Charpak, Georges and Sauli, Fabio},
  journal={Nuclear Instruments and Methods},
  volume={162},
  number={1-3},
  pages={405--428},
  year={1979},
  publisher={North-Holland}
}

@book{grupen2008particle,
  title={Particle detectors},
  author={Grupen, Claus and Shwartz, Boris},
  year={2008},
  publisher={Cambridge university press}
}

@article{tsyganov2008gas,
  title={Gas electron multiplying detectors for medical applications},
  author={Tsyganov, Edward and Antich, Peter and Parkey, Robert and Seliounine, Serguei and Golovatyuk, Viacheslav and Lobastov, Sergey and Zhezher, Valeriy and Buzulutskov, Alexei},
  journal={Nuclear Instruments and Methods in Physics Research Section A: Accelerators, Spectrometers, Detectors and Associated Equipment},
  volume={597},
  number={2-3},
  pages={257--265},
  year={2008},
  publisher={Elsevier}
}

@article{oed1995properties,
  title={Properties of micro-strip gas chambers (MSGC) and recent developments},
  author={Oed, A},
  journal={Nuclear Instruments and Methods in Physics Research Section A: Accelerators, Spectrometers, Detectors and Associated Equipment},
  volume={367},
  number={1-3},
  pages={34--40},
  year={1995},
  publisher={Elsevier}
}

@article{titov2013micro,
  title={Micro-pattern gaseous detector technologies and RD51 Collaboration},
  author={Titov, Maxim and Ropelewski, Leszek},
  journal={Modern Physics Letters A},
  volume={28},
  number={13},
  pages={1340022},
  year={2013},
  publisher={World Scientific}
}

@article{shekhtman2002micro,
  title={Micro-pattern gaseous detectors},
  author={Shekhtman, L},
  journal={Nuclear Instruments and Methods in Physics Research Section A: Accelerators, Spectrometers, Detectors and Associated Equipment},
  volume={494},
  number={1-3},
  pages={128--141},
  year={2002},
  publisher={Elsevier}
}

@article{sauli1997gem,
  title={GEM: A new concept for electron amplification in gas detectors},
  author={Sauli, Fabio},
  journal={Nuclear Instruments and Methods in Physics Research Section A: Accelerators, Spectrometers, Detectors and Associated Equipment},
  volume={386},
  number={2-3},
  pages={531--534},
  year={1997},
  publisher={Elsevier}
}

@article{sauli2004progress,
  title={Progress with the gas electron multiplier},
  author={Sauli, Fabio},
  journal={Nuclear Instruments and Methods in Physics Research Section A: Accelerators, Spectrometers, Detectors and Associated Equipment},
  volume={522},
  number={1-2},
  pages={93--98},
  year={2004},
  publisher={Elsevier}
}

@article{benlloch1998development,
  title={Development of the gas electron multiplier (GEM)},
  author={Benlloch, J and Bressan, Andrea and Buttner, C and Capeans, M and Gruwe, M and Hoch, M and Labb{\'e}, JC and Placci, A and Ropelewski, Leszek and Sauli, F and others},
  journal={IEEE Transactions on Nuclear Science},
  volume={45},
  number={3},
  pages={234--243},
  year={1998},
  publisher={IEEE}
}

@article{sauli2003development,
  title={Development and applications of gas electron multiplier detectors},
  author={Sauli, Fabio},
  journal={Nuclear Instruments and Methods in Physics Research Section A: Accelerators, Spectrometers, Detectors and Associated Equipment},
  volume={505},
  number={1-2},
  pages={195--198},
  year={2003},
  publisher={Elsevier}
}

@article{lyashenko2006advances,
  title={Advances in ion back-flow reduction in cascaded gaseous electron multipliers incorporating R-MHSP elements},
  author={Lyashenko, AV and Breskin, Amos and Chechik, Rachel and Veloso, JFCA and Dos Santos, JMF and Amaro, FD},
  journal={Journal of Instrumentation},
  volume={1},
  number={10},
  pages={P10004},
  year={2006},
  publisher={IOP Publishing}
}

@article{lyashenko2007further,
  title={Further progress in ion back-flow reduction with patterned gaseous hole-multipliers},
  author={Lyashenko, AV and Breskin, Amos and Chechik, Rachel and Veloso, JFCA and Dos Santos, JMF and Amaro, FD},
  journal={Journal of Instrumentation},
  volume={2},
  number={08},
  pages={P08004},
  year={2007},
  publisher={IOP Publishing}
}

@article{martyanov2014ansys,
  title={ANSYS Maxwell Software for electromagnetic field calculations},
  author={Martyanov, Andrey S and Neustroyev, Nikolay I},
  journal={Eastern European Scientific Journal},
  number={5},
  year={2014}
}

@book{thompson2017ansys,
  title={ANSYS mechanical APDL for finite element analysis},
  author={Thompson, Mary Kathryn and Thompson, John Martin},
  year={2017},
  publisher={Butterworth-Heinemann}
}

@manual{Garfieldppusemanual,
  title={Garfield++ User Guide},
  author={H. Schindler},
  year={2023},
url={http://garfieldpp.web.cern.ch/garfieldpp}
}

@article{Sarkar:2025tvu,
    author = "Sarkar, Amal and others",
    title = "{Dynamics of hot QCD matter 2024 {\textemdash} new facilities and instrumentation}",
    eprint = "2502.08187",
    archivePrefix = "arXiv",
    primaryClass = "hep-ex",
    doi = "10.1142/s021830132544001x",
    journal = "Int. J. Mod. Phys. E",
    volume = "34",
    number = "07",
    pages = "2544001",
    year = "2025"
}

@article{mondal2025optimizing,
  title={Optimizing Gas Electron Multiplier (GEM) geometry for enhanced gain, reduced ion backflow, and improved detector performance},
  author={Mondal, Md Kaosor Ali and Angiras, Poojan and Rana, Sachin and Sarkar, A},
  journal={Journal of Instrumentation},
  volume={20},
  number={04},
  pages={T04011},
  year={2025},
  publisher={IOP Publishing}
}

@article{Angiras:2025fcr,
    author = "Angiras, Poojan and Mondal, Md. Kaosor Ali and Sarkar, Amal",
    title = "{Modified GEM foil geometry for optimized gain and ion backflow reduction using ANSYS and Garfield++}",
    doi = "10.1016/j.jspc.2025.100240",
    journal = "J. Subatomic Part. Cosmol.",
    volume = "4",
    pages = "100240",
    year = "2025"
}

@misc{rana2026designoptimizationtriplegas,
      title={Design Optimization of Triple Gas Electron Multiplier for Superior Gain and Reduced Ion Backflow}, 
      author={Sachin Rana and Md. Kaosor Ali Mondal and Poojan Angiras and Amal Sarkar},
      year={2026},
      eprint={2601.12553},
      archivePrefix={arXiv},
      primaryClass={physics.ins-det},
      url={https://arxiv.org/abs/2601.12553},
}

@article{sauli2016gem,
  title={The gas electron multiplier ({GEM}): Operating principles and applications},
  author={Sauli, Fabio},
  journal={Nuclear Instruments and Methods in Physics Research Section A: Accelerators, Spectrometers, Detectors and Associated Equipment},
  volume={805},
  pages={2--24},
  year={2016},
  doi={10.1016/j.nima.2015.07.060},
  publisher={Elsevier}
}

@article{bachmann1999charge,
  title={Charge amplification and transfer processes in the gas electron multiplier},
  author={Bachmann, S and Bressan, A and Ropelewski, L and Sauli, F and Sharma, A and M{\"o}rmann, D},
  journal={Nuclear Instruments and Methods in Physics Research Section A: Accelerators, Spectrometers, Detectors and Associated Equipment},
  volume={438},
  number={2-3},
  pages={376--408},
  year={1999},
  doi={10.1016/S0168-9002(99)00820-7},
  publisher={Elsevier}
}

@article{sauli2003electron,
  title={Electron collection and ion feedback in {GEM}-based detectors},
  author={Sauli, Fabio and Kappler, Steffen and Ropelewski, Leszek},
  journal={IEEE Transactions on Nuclear Science},
  volume={50},
  number={4},
  pages={803--808},
  year={2003},
  doi={10.1109/TNS.2003.814543},
  publisher={IEEE}
}

@article{bondar2003study,
  title={Study of ion feedback in multi-{GEM} structures},
  author={Bondar, A and Buzulutskov, A and Shekhtman, L and Vasiljev, A},
  journal={Nuclear Instruments and Methods in Physics Research Section A: Accelerators, Spectrometers, Detectors and Associated Equipment},
  volume={496},
  number={2-3},
  pages={325--332},
  year={2003},
  doi={10.1016/S0168-9002(02)01763-1},
  publisher={Elsevier}
}

@article{ketzer2004performance,
  title={Performance of triple {GEM} tracking detectors in the {COMPASS} experiment},
  author={Ketzer, B and Weitzel, Q and Paul, S and Sauli, F and Ropelewski, L},
  journal={Nuclear Instruments and Methods in Physics Research Section A: Accelerators, Spectrometers, Detectors and Associated Equipment},
  volume={535},
  number={1-2},
  pages={314--318},
  year={2004},
  doi={10.1016/S0168-9002(04)01668-7},
  publisher={Elsevier}
}

@techreport{ALICE:2013TPCTDR,
  title={Technical Design Report for the Upgrade of the {ALICE} Time Projection Chamber},
  author={{ALICE Collaboration}},
  institution={CERN},
  number={CERN-LHCC-2013-020, ALICE-TDR-016},
  address={Geneva},
  year={2013}
}

@techreport{ALICE:2015TPCTDRadd,
  title={Addendum to the Technical Design Report for the Upgrade of the {ALICE} Time Projection Chamber},
  author={{ALICE Collaboration}},
  institution={CERN},
  number={CERN-LHCC-2015-002, ALICE-TDR-016-ADD-1},
  address={Geneva},
  year={2015}
}

@article{ALICETPC:2021upgrade,
  title={The upgrade of the {ALICE} {TPC} with {GEMs} and continuous readout},
  author={{ALICE TPC Collaboration}},
  journal={Journal of Instrumentation},
  volume={16},
  number={03},
  pages={P03022},
  year={2021},
  doi={10.1088/1748-0221/16/03/P03022},
  eprint={2012.09518},
  archivePrefix={arXiv},
  publisher={IOP Publishing}
}

@techreport{CMS:2015GEMTDR,
  title={{CMS} Technical Design Report for the Muon Endcap {GEM} Upgrade},
  author={Colaleo, A and Safonov, A and Sharma, A and Tytgat, M},
  institution={CERN},
  number={CERN-LHCC-2015-012, CMS-TDR-013},
  address={Geneva},
  year={2015}
}

@article{ball2014ion,
  title={Ion backflow studies for the {ALICE} {TPC} upgrade with {GEMs}},
  author={Ball, M and Eckstein, K and Gunji, T},
  journal={Journal of Instrumentation},
  volume={9},
  number={04},
  pages={C04025},
  year={2014},
  doi={10.1088/1748-0221/9/04/C04025},
  publisher={IOP Publishing}
}

@article{boehmer2013simulation,
  title={Simulation of space-charge effects in an ungated {GEM}-based {TPC}},
  author={B{\"o}hmer, F V and Ball, M and D{\o}rheim, S and H{\"o}ppner, C and Ketzer, B and Konorov, I and Neubert, S and Paul, S and Rauch, J and Vandenbroucke, M},
  journal={Nuclear Instruments and Methods in Physics Research Section A: Accelerators, Spectrometers, Detectors and Associated Equipment},
  volume={719},
  pages={101--108},
  year={2013},
  doi={10.1016/j.nima.2013.04.020},
  eprint={1209.0482},
  archivePrefix={arXiv},
  publisher={Elsevier}
}

@article{natal2018ion,
  title={Ion backflow studies with a triple-{GEM} stack with increasing hole pitch},
  author={Natal da Luz, H and Bhattacharya, P and Filho, L A S and Fran{\c{c}}a, L E F M},
  journal={Journal of Instrumentation},
  volume={13},
  number={07},
  pages={P07025},
  year={2018},
  doi={10.1088/1748-0221/13/07/P07025},
  eprint={1803.09382},
  archivePrefix={arXiv},
  publisher={IOP Publishing}
}

@article{bhattacharya2017simulation,
  title={{3D} simulation of electron and ion transmission of {GEM}-based detectors},
  author={Bhattacharya, Purba and Mohanty, Bedangadas and Mukhopadhyay, Supratik and Majumdar, Nayana and Natal da Luz, Hugo},
  journal={Nuclear Instruments and Methods in Physics Research Section A: Accelerators, Spectrometers, Detectors and Associated Equipment},
  volume={870},
  pages={64--72},
  year={2017},
  doi={10.1016/j.nima.2017.06.054},
  eprint={1710.00607},
  archivePrefix={arXiv},
  publisher={Elsevier}
}

@article{patra2018characteristic,
  title={Characteristic study of a quadruple {GEM} detector and its comparison with a triple {GEM} detector},
  author={Patra, Rajendra Nath and Singaraju, Rama Narayan and Biswas, Saikat and Viyogi, Yogendra P and Nayak, Tapan K},
  journal={Nuclear Instruments and Methods in Physics Research Section A: Accelerators, Spectrometers, Detectors and Associated Equipment},
  volume={906},
  pages={37--42},
  year={2018},
  doi={10.1016/j.nima.2018.08.023},
  eprint={1810.03316},
  archivePrefix={arXiv},
  publisher={Elsevier}
}

@article{greene2022effective,
  title={Effective gain and ion back flow study of triple and quadruple {GEM} detector},
  author={Greene, Senta V and Velkovska, Julia and Blankenship, Brandon and Reynolds, Michael Z and Tarafdar, Sourav},
  journal={Journal of Instrumentation},
  volume={17},
  number={12},
  pages={T12004},
  year={2022},
  doi={10.1088/1748-0221/17/12/T12004},
  eprint={2011.14568},
  archivePrefix={arXiv},
  publisher={IOP Publishing}
}

@article{alfonsi2012simulation,
  title={Simulation of the dielectric charging-up effect in a {GEM} detector},
  author={Alfonsi, M and Croci, G and Duarte Pinto, S and Rocco, E and Ropelewski, L and Sauli, F and Veenhof, R and Villa, M},
  journal={Nuclear Instruments and Methods in Physics Research Section A: Accelerators, Spectrometers, Detectors and Associated Equipment},
  volume={671},
  pages={6--9},
  year={2012},
  doi={10.1016/j.nima.2011.12.059},
  publisher={Elsevier}
}

@article{correia2014dynamic,
  title={A dynamic method for charging-up calculations: the case of {GEM}},
  author={Correia, P M M and Oliveira, C A B and Azevedo, C D R and Silva, A L M and Veenhof, R and Nemallapudi, M V and Veloso, J F C A},
  journal={Journal of Instrumentation},
  volume={9},
  number={07},
  pages={P07025},
  year={2014},
  doi={10.1088/1748-0221/9/07/P07025},
  eprint={1401.4009},
  archivePrefix={arXiv},
  publisher={IOP Publishing}
}

@article{villa2011progress,
  title={Progress on large area {GEMs}},
  author={Villa, Marco and Duarte Pinto, Serge and Alfonsi, Matteo and Brock, Ian and Croci, Gabriele and David, Eric and de Oliveira, Rui and Ropelewski, Leszek and Taureg, Hans and van Stenis, Miranda},
  journal={Nuclear Instruments and Methods in Physics Research Section A: Accelerators, Spectrometers, Detectors and Associated Equipment},
  volume={628},
  number={1},
  pages={182--186},
  year={2011},
  doi={10.1016/j.nima.2010.06.312},
  eprint={1007.1131},
  archivePrefix={arXiv},
  publisher={Elsevier}
}

@article{biagi1999monte,
  title={Monte Carlo simulation of electron drift and diffusion in counting gases under the influence of electric and magnetic fields},
  author={Biagi, S F},
  journal={Nuclear Instruments and Methods in Physics Research Section A: Accelerators, Spectrometers, Detectors and Associated Equipment},
  volume={421},
  number={1-2},
  pages={234--240},
  year={1999},
  doi={10.1016/S0168-9002(98)01233-9},
  publisher={Elsevier}
}

@article{veenhof1998garfield,
  title={Garfield, recent developments},
  author={Veenhof, Rob},
  journal={Nuclear Instruments and Methods in Physics Research Section A: Accelerators, Spectrometers, Detectors and Associated Equipment},
  volume={419},
  number={2-3},
  pages={726--730},
  year={1998},
  doi={10.1016/S0168-9002(98)00851-1},
  publisher={Elsevier}
}

\end{document}